\documentstyle[aps]{revtex}
\begin{document}

\title{From Davydov solitons to decoherence-free subspaces: \\ 
self-consistent propagation of coherent-product states}
\author{S. Gheorghiu-Svirschevski\footnotemark[1]\footnotetext{e-mail: hnmg@soa.com}}
\address{1087 Beacon St., Suite 301, Newton, Massachusetts 02459}
\date{\today}
\maketitle
\begin{abstract}
The self-consistent propagation of generalized $D_{1}$ [coherent-product] states and of a class of gaussian density matrix generalizations is examined, at both zero and finite-temperature, for arbitrary interactions between the localized lattice (electronic or vibronic) excitations and the phonon modes. It is shown that in all legitimate cases, the evolution of $D_{1}$ states reduces to the disentangled evolution of the component $D_{2}$ states. The self-consistency conditions for the latter amount to conditions for decoherence-free propagation, which complement the $D_{2}$ Davydov soliton equations in such a way as to lift the nonlinearity of the evolution for the on-site degrees of freedom. Although it cannot support Davydov solitons, the coherent-product ansatz does provide a wide class of exact density-matrix solutions for the joint evolution of the lattice and phonon bath in compatible systems. Included are solutions for initial states given as a product of a [largely arbitrary] lattice state and a thermal equilibrium state of the phonons. It is also shown that external pumping can produce self-consistent Frohlich-like effects. A few sample cases of coherent, albeit not solitonic, propagation are briefly discussed. 
\end{abstract}


\section{Introduction}
\label{sec1}
Davydov's model of soliton propagation in molecular chains is not rigorous \cite{invalid,brown}, but its transparency and physical appeal continue to encourage intense work regarding its utility as a practical approximation, at both zero and finite-temperature \cite{approx}. Alternatively, it is commonly hoped that more elaborate Hamiltonians and/or wave functions may render the model exact or improve its accuracy, and various proposed refinements have attempted to do so \cite{improved}. From the latter point of view, it is quite intriguing that the conventional approach to the problem stops short of probing the self-consistency conditions for the wave function ansatz. Indeed, the Davydov model relies on two fundamental assumptions: i) A $D_1$ [$D_2$] ansatz state is a solution of the Schroedinger equation for the Davydov Hamiltonian. ii) The time-dependence of the ansatz state can be obtained by treating the ansatz parameters as canonical variables in a Hamilton's functional given by the average of the Davydov Hamiltonian on the ansatz state. The second conjecture finds a self-consistent foundation in the variational principle of least action \cite{variational}, while the first is suspended, for practical purposes, and the state ansatz is cast as a variational trial ansatz. But the ansatz state is likely to be a good [dynamical] trial state in those situationsthat are 'close', in some suitable, perturbative sense, to a self-consistent model for which this state is an exact solution. Therefore, one can start, conceivably, by questioning what particular circumstances allow an entangled superposition of coherent or thermal-coherent [Gaussian] phonon states to preserve its cohesion in time. 

The present paper aims to address this problem in  a general setting, at both zero and finite-temperature. The outcome, which can be viewed, eventually, as a generalization of two previous theorems, due to Brown \cite{brown}, on the validity of the standard $D_2$ and $D_1$ states, has the unexpected effect to place the Davydov ansatz problem in a fresh perspective. Our  starting point is a generic, not necessarily unidimensional, lattice of $N$ interacting monomers, each coupled in turn to a common phonon [bosonic] band. No {\it a priori} assumptions are made as to the nature of the lattice [monomer] degrees of freedom [excitonic or vibronic], of the phonon modes [acoustic/longitudinal, optical/dispersionless, etc.], or of the site-to-site and lattice-phonons interactions. Not surprisingly, however, the very nature of the coherent states, as states specific to harmonic systems, limits the type of appropriate lattice-phonon interactions to bilinear terms in the phonon degrees of freedom at zero temperature, and to only linear terms at finite-temperature. This notwithstanding, the critical constraints concern the accompanying, entangled lattice states and the lattice contribution to the lattice-phonon interaction. Somewhat contrary to the widely held view that the coherent phonons should drive the (self-trapped) lattice configuration, it turns out that phonon coherence is also essentially conditioned by the nature of the entangled lattice states. We find that self-consistency requires, both at zero and at finite-temperature, that the dynamics of  $D_{1}$ superpositions be reduced, remarkably, to the disentangled propagation of the $D_{2}$ components. Therefore no self-consistent  $D_{1}$ model can generate soliton equations coupling the separate $D_{2}$ states. Further, the lattice state in any $D_{2}$ product must satisfy constraints leading to [slightly modified] decoherence-free evolution \cite{decoherence1}. As a result, the self-consistent lattice states propagate unitarily and all potential nonlinearities in their effective equation of motion cancel identically. Hence, no $D_{2}$ model based on a linear decomposition of the lattice state can generate Davydov-type soliton equations for the associated [linearly independent] lattice amplitudes. 

A stronger and quite peculiar result concerns systems with lattice-phonon couplings linear in the phonon coordinates. First, only such systems can support generalized, Gaussian $D_{2}$ states at finite-temperatures. Second, the corresponding self-consistency conditions for lattice states become precisely the constraints that typically define decoherence-free subspaces [DFS], as recently considered in connection with noiseless quantum computation \cite{decoherence1}, \cite{decoherence2}, and the effective Hamiltonian reduces to the unperturbed lattice Hamiltonian [up to a phonon-modulated energy shift, which can be transferred onto the phonon state]. When the coupling is time independent, the associated phonon displacements perform simple harmonic oscillations around a displaced equilibrium position. One is so led to the conclusion that finite-temperature coherent-product propagation is possible if and only if the lattice-phonon coupling is linear in the phonon coordinates and decoherence-free subspaces exist for the lattice subsystem. 

From the point of view of the soliton problem, this outcome relates to a rather self-evident idea in the context of decoherence-free propagation. That is, if an unperturbed lattice can support soliton states that evolve entirely within a DFS, such solitons will propagate unperturbed [at arbitrary temperatures], regardless of the state of the environment. Obviously, coherent-product states loose any special significance in such a solution, and the problem falls outside the scope of the present paper.

Nevertheless, the self-consistent coherent-product ansatz demonstrates a sufficient number of notable features to remain attractive by itself. For instance, let us recall that null phonon displacements place the bath in a thermal state, hence initial null displacements reduce any initial Gaussian $D_{1}$ state to the ubiquitous product of a lattice state and a phonon thermal state. Since the initial lattice state can be an arbitrary distribution on a direct sum of [orthogonal] DFS, we obtain the following corollary, with reference to the theory of decoherence-free propagation. 

{\it Let a system interact with a boson bath through a coupling linear in boson coordinates. Any distribution on a direct sum of system DFS develops, when brought in contact with the bath in a thermal state, into a [strongly entangled] Gaussian  $D_{1}$ state. }

Because the expression of the  $D_{1}$ state is exactly known, the development of the system-bath entanglement in such a model can be monitored precisely. The result is a remarkable counterexample to the standard picture of relaxation through thermal contact. Regardless of any intrinsic characteristics, the bath is forced into a coherent, nonequilibrium state, while the DFS components of the system state evolve in an unperturbed manner. This confirms earlier claims that decoherence may produce in fact coherent states \cite{coh-decoh1}, and therefore caution should be exerted in assuming that a heat bath remains at all times in thermal equilibrium \cite{coh-decoh2}. It also corroborates the similar conclusion of a recent study \cite{Venugopalan} of decoherence in a quantum system interacting with a [macroscopic] measuring apparatus. When the correlations between distinct DFS vanish in time, the system is driven toward a statistical superposition of decoherence-free, pointer states and  we retrieve a typical example of environment-induced superselection \cite{zurek}. But in another interesting limit, which arises under time independent interactions, the bath modes can be prepared such that the initial distortions match the displaced equilibrium positions for the coherent oscillations. In this situation, the bath remains in a stationary nonequilibrium state, while the system evolves unitarily, according to the unperturbed dynamics [up to time independent DFS energy shifts], irrespective of any entanglement with the bath. Leaving aside questions of stability, such a state provides an apparent counterexample to environment-induced superselection, within the same physical system. 

Along a different line of inquiry, proper external pumping can be used to promote decoherence-free propagation in systems that otherwise may not have the necessary symmetries. We find, incidentally, that such a process can be accompanied by a Frohlich-like effect on the bath modes. That is, pumping at a frequency attuned to the lattice subsystem may result in a macroscopic displacement of a bath mode of a different frequency, while other reservoir modes remain in thermal equilibrium. 

The paper is organized as follows. We begin with the familiar pure-state case in Sec. II. Section IV develops the density-matrix generalization, which includes Davydov's thermal ansatz and allows a straightforward approach to the finite-temperature problem. To this end, we employ a formal framework, outlined in Sec. III, based on the 'square-root' decomposition  of the density-matrix and the concepts of thermal vacuum and thermal Fock space introduced in thermofield dynamics [TFD] \cite{thermo}. This formalism can be regarded in effect as a version of TFD without auxiliary systems. Our choice is motivated by the notable technical advantage that pure state calculations can be effortlessly transcribed into the density-matrix domain, with a minimum of adjustments. Section V examines and discusses some popular versions of Davydov's model, alongside with sample self-consistent examples, including a case with time-dependent interaction (external pumping). A summary and concluding remarks are provided in Sec. VI.

\section{Self-consistent dynamics of $D_{1}$ states at $T=0$}
\label{Sec2}

Let the lattice-phonon Hamiltonian be

\begin{equation}
\label{eq1}
H = H_{lat}  + H_{ph}^0  + W\;,
\end{equation}

\noindent
where

\begin{equation}
\label{eq2}
H_{lat}  = \sum\limits_{n=0}^{N-1} {\varepsilon _n c_n^{\dagger}  c_n }  + V_{lat}
\end{equation}

\noindent describes the $N$-site lattice, with  $V_{lat}$ subsuming all hoping interactions, 

\begin{equation}
\label{eq3}
H_{ph}^0  = \sum\limits_q {\hbar \omega _{q} b_{q}^{\dagger}  b_{q} }
\end{equation}

\noindent
is the bare phonon Hamiltonian,  and $W$ accounts for any lattice-phonon and/or anharmonic phonon interactions. 
Let us search for self-sustained soliton states $\left| \Psi  \right\rangle $ in the slightly generalized Davydov $D_1$ ansatz 

\begin{equation}
\label{eq4}
\left| \Psi  \right\rangle  = \sum\limits_\alpha  {\left| \alpha  \right\rangle } \left| {\beta _\alpha  } \right\rangle \;,
\end{equation}

\noindent 
where the $\left| \alpha  \right\rangle$'s are orthogonal, but not normalized, lattice (vibronic) states, $\left\langle {\alpha '|\left. \alpha  \right\rangle } \right. \sim \delta _{\alpha '\alpha } $, $\sum\limits_\alpha  {\left\langle {\alpha } \mathrel{\left | {\vphantom {\alpha  \alpha }} \right. \kern-\nulldelimiterspace} {\alpha } \right\rangle  = 1} $, and the $\left| {\beta _\alpha  } \right\rangle$'s are coherent phonon states, 

\begin{equation}
\label{eq5}
\left| {\beta _\alpha  } \right\rangle  = \exp \left[ {\sum\limits_q {\left( {\beta _{q\alpha } b_q^{\dagger}   - \beta _{q\alpha }^* b_q } \right)} } \right]{\rm  }\left| {\rm 0} \right\rangle _{ph} \;.
\end{equation}

\noindent
Here the $\left| \alpha  \right\rangle$'s are allowed to contain local excitations in different numbers and are not necessarily confined to the same site. The $\left| \Psi  \right\rangle $ ansatz above is, thus, general enough to cover both single-quantum and multiquanta solitons, as well as their eventual superpositions. The set of occupied states $\left\{ {\left| \alpha  \right\rangle } \right\}_{\left\langle {\alpha }\mathrel{\left | {\vphantom {\alpha  \Psi }} \right. \kern-\nulldelimiterspace}{\Psi } \right\rangle  \ne 0} $ is to be regarded as embedded into an orthogonal basis $\left\{ {\left| \alpha  \right\rangle }\right\}$ of lattice states; the unoccupied $\left| \alpha  \right\rangle$'s, $\left\langle {\alpha }\mathrel{\left | {\vphantom {\alpha  \Psi }}\right. \kern-\nulldelimiterspace}{\Psi } \right\rangle  = 0$, may be assigned by default $\left| {\beta _\alpha  } \right\rangle  = \left| 0 \right\rangle _{ph} $. For each displaced phonon vacuum $\left| {\beta _\alpha  } \right\rangle $, let us construct the corresponding orthonormal Fock basis  

\begin{equation}
\label{eq6}
\left| { \ldots n_{q,\alpha }  \ldots } \right\rangle  = \prod\limits_q {\frac{1}{{\sqrt {n_q !} }}\left( {b_q^{\dagger}   - \beta _{q\alpha }^ * } \right)^{n_q } \left| {\beta _\alpha  } \right\rangle } \;,
\end{equation}

\noindent
where $\left| { \ldots n_{q,\alpha }  \ldots } \right\rangle $ carries $n_{q,\alpha } $  displaced quanta of mode $q$ relative to $\left| {\beta _\alpha  } \right\rangle $. In particular, $\left| {1_{q,\alpha } } \right\rangle  = \left( {b_q^{\dagger}   - \beta _{q\alpha }^ *  } \right)\left| {\beta _\alpha  } \right\rangle $ denotes the first excited state of the displaced mode $q$ relative to the displaced vacuum $\left| {\beta _\alpha  } \right\rangle $. The set $\left\{ {\left| \alpha  \right\rangle \left| { \ldots n_{q,\alpha }  \ldots } \right\rangle } \right\}_{\alpha ,{\rm  }\sum\limits_q {n_{q,\alpha } }  \ge 0} $
 obviously provides an orthogonal basis for the overall lattice-phonon system.

Extending the idea originally applied by Brown in Ref.\cite{brown}, the conditions under which the state $\left| \Psi  \right\rangle $ of Eq. (\ref{eq4}) is compatible with the dynamics driven by $H$ will be derived from the expansion of the corresponding Schroedinger equation

\begin{equation}
\label{eq7}
i\hbar \frac{d}{{dt}}\left| \Psi  \right\rangle  = H\left| \Psi  \right\rangle
\end{equation}

\noindent 
in this displaced  phonon basis. Indeed, substitution of the explicit form of $\left| \Psi  \right\rangle $, followed by some straightforward algebraic manipulation, leads to 

\[ i\hbar \left[ {\sum\limits_\alpha  {\left| {\dot \alpha } \right\rangle \left| {\beta _\alpha  } \right\rangle }  + \frac{1}{2}\sum\limits_{\alpha ,q} {\left( {\dot \beta _{q\alpha } \beta _{q\alpha }^ *   - \beta _{q\alpha } \dot \beta _{q\alpha }^ *  } \right)\left| \alpha  \right\rangle \left| {\beta _\alpha  } \right\rangle }  + \sum\limits_{\alpha ,q} {\dot \beta _{q\alpha } \left| \alpha  \right\rangle \left| {1_{q,\alpha } } \right\rangle } } \right] =    \]

\begin{equation}
\label{eq8}
= \sum\limits_\alpha  {\left( {H_{ex}  + W + \sum\limits_q {\hbar \omega _q \left| {\beta _{q\alpha } } \right|} ^2 } \right)\left| \alpha  \right\rangle \left| {\beta _\alpha  } \right\rangle }  + \sum\limits_{\alpha ,q} {\hbar \omega _q \beta _{q\alpha } \left| \alpha  \right\rangle \left| {1_{q,\alpha } } \right\rangle }  \;.
\end{equation}

\noindent
Since the time derivative of $\left| \Psi  \right\rangle $ on the left-hand side of Eq. (\ref{eq8}) carries only terms in displaced vacuums and their first excited states, it is foreseeable that the main constraints will emerge from the requirement that higher order contributions on the right-hand side vanish identically. As previously hinted, we can expect the interaction $ $ to be limited to bilinear terms in the phonon coordinates. But it will also become apparent that the form of the effective equation of motion and the self-consistency conditions for the lattice states $\left| \alpha  \right\rangle $ are shaped by this same requirement. So consider first the contraction of both sides of  Eq. (\ref{eq8}) with states  $\left| {\alpha '} \right\rangle \left| { \ldots n_{q,\alpha '}  \ldots } \right\rangle $ carrying more than one phonon quantum, $\sum\limits_q {n_{q} }  > 1$. After conveniently expanding the contribution in $W$ along displaced phonon bases and rearranging the terms, we are left with

\[ \sum\limits_{\alpha  \ne \alpha '} {\left[ {\left\langle {\alpha '\left| {H_{ex} } \right|\left. \alpha  \right\rangle } \right. + \left\langle {\alpha '} \right.\left| {\left\langle {\beta _\alpha  \left| W \right|\left. {\beta _\alpha  } \right\rangle } \right.} \right|\left. \alpha  \right\rangle  - i\hbar \left\langle {{\alpha '}}\mathrel{\left | {\vphantom {{\alpha '} {\dot \alpha }}} \right. \kern-\nulldelimiterspace}{{\dot \alpha }} \right\rangle } \right]\left\langle {{ \ldots n_{q',\alpha '}  \ldots }} \mathrel{\left | {\vphantom {{ \ldots n_{q',\alpha '}  \ldots } {\beta _\alpha  }}} \right. \kern-\nulldelimiterspace}{{\beta _\alpha  }} \right\rangle } + \]

\[ + \sum\limits_{\alpha  \ne \alpha ',{\rm  q}} {\left\langle {\alpha '} \right.\left| {\left\langle {1_{q,\alpha } \left| W \right|\left. {\beta _\alpha  } \right\rangle } \right.} \right|\left. \alpha  \right\rangle \left\langle {{ \ldots n_{q',\alpha '}  \ldots }} \mathrel{\left | {\vphantom {{ \ldots n_{q',\alpha '}  \ldots }{1_{q,\alpha } }}} \right. \kern-\nulldelimiterspace}{{1_{q,\alpha } }} \right\rangle }  +\]

\begin{equation}
\label{eq9}
+  \sum\limits_\alpha  {\sum\limits_{\scriptstyle {\rm    }\left\{ {{\rm m}_{\rm q} } \right\} \hfill \atop  \scriptstyle \sum\limits_q {m_q }  > 1 \hfill} {\left\langle {\alpha '} \right.\left| {\left\langle { \ldots m_{q,\alpha }  \ldots \left| W \right|\left. {\beta _\alpha  } \right\rangle } \right.} \right|\left. \alpha  \right\rangle \left\langle {{ \ldots n_{q',\alpha '} \ldots }} \mathrel{\left | {\vphantom {{ \ldots n_{q',\alpha '}  \ldots } { \ldots m_{q,\alpha }  \ldots }}}\right. \kern-\nulldelimiterspace}{{ \ldots m_{q,\alpha }  \ldots}} \right\rangle } }  = 0  \;.   
\end{equation}

\noindent
Under the reasonable assumption that the form of the interaction $W$ limits the last sum above to a finite number of terms, a nontrivial solution will be compatible with the infinite number of constraints (\ref{eq9}) if and only if all quantities multiplying the nonvanishing overlap factors $\left\langle {{ \ldots n_{q',\alpha '}  \ldots }}
 \mathrel{\left | {\vphantom {{ \ldots n_{q',\alpha '}  \ldots } { \ldots k_{q,\alpha } \ldots }}} \right. \kern-\nulldelimiterspace}{{ \ldots k_{q,\alpha }  \ldots }} \right\rangle $ cancel identically. In particular, the terms under the last sum in Eq. (\ref{eq9}) imply

\begin{equation}
\label{eq10}
\left\langle { \ldots m_{q,\alpha }  \ldots \left| W \right|\left. {\beta _\alpha  } \right\rangle } \right. = 0, \;\;\sum\limits_q {m_{q,\alpha } }  > 1.
\end{equation}

\noindent
Hence $W$ can only contain terms in $b_q ,{\rm  }b_q^{\dagger}  $ and $b_q^{\dagger}  b_{q'}$, and must be of the form

\begin{equation}
\label{eq11}
W = \sum\limits_q {\left( {b_q^{\dagger}  w_q  + b_q w_q^{\dagger}  } \right) + \sum\limits_{q,q'} {b_q^{\dagger}  b_{q'} u_{qq'} } } \;,
\end{equation}

\noindent
where the operators $w_{q}$ and $u_{qq'}$ act on on-site degrees of freedom only, as (polynomial) functions of  $c_{n}$, $c_{n}^{+}$, and the hermiticity of $W$ requires $u_{q'q}^{\dagger}   = u_{qq'} $. Further, from the second sum in Eq. (\ref{eq9}) it can be inferred that  

\begin{equation}
\label{eq12}
\left\langle {\alpha '} \right.\left| {\left\langle {1_{q,\alpha } \left| W \right|\left. {\beta _\alpha  } \right\rangle } \right.} \right|\left. \alpha  \right\rangle  = 0 \; \;\forall q,{\rm  }\alpha  \ne \alpha '{\rm  }\;.
\end{equation}

\noindent 
But since $\left| {\alpha '} \right\rangle $ spans an entire lattice basis, the above constraints imply in fact that the occupied states $\left| \alpha  \right\rangle $ must satisfy eigenvalue equations of the form

\begin{equation}
\label{eq13}
G_{q,\alpha } \left| \alpha  \right\rangle  = \gamma _{q,\alpha } \left| \alpha  \right\rangle ,\;\;\forall \alpha ,q\;,
\end{equation}

\noindent
for $G_{q,\alpha }  = \left\langle {1_{q,\alpha } } \right.\left| W \right|\left. {\beta _\alpha  } \right\rangle $. In the ansatz (\ref{eq11}) for $W$, the operators $G_{q,\alpha }$ read

\begin{equation}
\label{eq14}
G_{q,\alpha }  = w_q  + \sum\limits_{q'} {\beta _{q'\alpha } u_{qq'} } \;.
\end{equation}

\noindent
Similarly, from the first sum in Eq. (\ref{eq9}) it follows that 

\begin{equation}
\label{eq15}
\left\langle {\alpha '\left| {H_{ex} } \right|\left. \alpha  \right\rangle } \right. + \left\langle {\alpha '} \right.\left| {\left\langle {\beta _\alpha  \left| W \right|\left. {\beta _\alpha  } \right\rangle } \right.} \right|\left. \alpha  \right\rangle  - i\hbar \left\langle {{\alpha '}}
 \mathrel{\left | {\vphantom {{\alpha '} {\dot \alpha }}}
 \right. \kern-\nulldelimiterspace}
 {{\dot \alpha }} \right\rangle  = 0{\rm   }\;,\;\;\forall \alpha ' \ne \alpha \;,
\end{equation}

\noindent
which shows that the occupied $\left| \alpha  \right\rangle$'s, $\left\langle {\alpha} \mathrel{\left | {\vphantom {\alpha  \Psi }}\right. \kern-\nulldelimiterspace}{\Psi } \right\rangle  \ne 0$, should evolve according to an effective Shroedinger equation 

\begin{equation}
\label{eq16}
i\hbar \left| {\dot \alpha } \right\rangle  = \left[ {H_{ex}  + \left\langle {\beta _\alpha  } \right.\left| W \right|\left. {\beta _\alpha  } \right\rangle  + \Omega _\alpha  } \right]\left| \alpha  \right\rangle \;,
\end{equation}

\noindent
with $\Omega _\alpha  $ a scalar functional, and  

\[\; \;\;\left\langle {\beta _\alpha  } \right|W\left| {\beta _\alpha  } \right\rangle  = \sum\limits_q {\left( {\beta _{q\alpha }^* w_q  + \beta _{q\alpha } w_q^{\dagger}  } \right) + \sum\limits_{q,q'} {\beta _{q\alpha }^* \beta _{q'\alpha } u_{qq'} } }=  \]

\begin{equation}
\label{eq17}
= \sum\limits_q {\left( {\beta _{q\alpha } w_q^{\dagger} +  \beta _{q\alpha }^ *  G_{q\alpha } } \right)}= \sum\limits_q {\left( {\beta _{q\alpha }^*  w_q  + \beta _{q\alpha } G_{q\alpha }^{\dagger}  } \right)}  \;.
\end{equation}
 
It is already evident, from Eqs.(\ref{eq13}) and (\ref{eq16}), that the evolution of distinct lattice states $\left| \alpha  \right\rangle $ is reciprocally decoupled, unless the eigenvalues $\mu _{q,\alpha } $ exhibit a dependence on some $\beta _{q\alpha '} $ with $\alpha ' \ne \alpha $. It will be seen shortly that this is not the case. At the same time, it can be recognized that the eigenvalue equations (\ref{eq13}) implement an effective decoupling of the lattice-phonon interaction, very much in the manner of the effective decoupling responsible for decoherence-free subspaces \cite{decoherence1}. The latter process requires that, for a lattice-phonons interaction of the general form $\sum\limits_k {W_k^{(lat)} U_k^{(ph)} } $, the lattice state $\left| {\Psi _{lat} } \right\rangle $
 be such that $W_k^{(lat)} \left| {\Psi _{lat} } \right\rangle  = \mu _k \left| {\Psi _{lat} } \right\rangle $ for all k, and at all times. A subspace of lattice states complying with these constraints is termed a decoherence-free subspace (DFS). Obviously, states belonging to a lattice DFS are completely decoupled from the phonon dynamics and evolve unperturbed. A closer examination of our conditions (\ref{eq13}) shows that, in fact, the present process differs from a true DFS selection simply by the assumption of a coherent ansatz for the state of the phonon modes. When the interaction in Eq. (\ref{eq11}) is applied to any product $\left| \alpha  \right\rangle \left| {\beta _\alpha  } \right\rangle $, the result reads 

\[ W\left| \alpha  \right\rangle \left| {\beta _\alpha  } \right\rangle  = \left[ {\sum\limits_q {b_q^{\dagger}  \left( {w_q  + \sum\limits_{q'} {\beta _{q'\alpha } u_{qq'} } } \right)\left| \alpha  \right\rangle } } \right]\left| {\beta _\alpha  } \right\rangle  + \left[ {\sum\limits_q {\beta _{q\alpha } w_q^{\dagger}  } \left| \alpha  \right\rangle } \right]\left| {\beta _\alpha  } \right\rangle\;, 
\]

\noindent and it is immediate that condition (\ref{eq13}) is just the proper condition for the decoupling of the first term on the right-hand side. It must be kept in mind, however, that we have arrived at this result without any assumptions on the form of $W$ or on the separability of the $D_2$ terms at the outset. 

The details in the equation of motion (\ref{eq16}) and the self-consistency constraints (\ref{eq13}) will follow, as can be anticipated, from a balance of the left-hand side of Eq. (\ref{eq8}) with similar, nonvanishing terms on its right-hand side. Indeed, let us first contract Eq. (\ref{eq8}) with states $\left| {\alpha '} \right\rangle \left| {\beta _{\alpha '} } \right\rangle $. Taking into account conditions (\ref{eq15}) above, we are led to  
  
\begin{equation}
\label{eq18}
i\hbar \left\langle {\alpha } \mathrel{\left | {\vphantom {\alpha  {\dot \alpha }}}
 \right. \kern-\nulldelimiterspace} {{\dot \alpha }} \right\rangle  = \left\langle \alpha  \right.\left| {\left( {H_{ex}  + \left\langle {\beta _\alpha  \left| W \right|\left. {\beta_\alpha} \right\rangle } \right.} \right)} \right|\left. \alpha  \right\rangle  + \sum\limits_q {\left[ {\hbar \omega _q \left| {\beta _{q\alpha } } \right|^2  - \frac{{i\hbar }}{2}\left( {\dot \beta _{q\alpha } \beta _{q\alpha }^ *   - \beta _{q\alpha } \dot \beta _{q\alpha }^ *  } \right)} \right]} \left\langle {\alpha }\mathrel{\left | {\vphantom {\alpha  \alpha }}\right. \kern-\nulldelimiterspace}{\alpha }\right\rangle \;,
\end{equation}

\noindent 
which, in conjunction with Eqs.(\ref{eq16}), identifies

\begin{equation}
\label{eq19}
\Omega _\alpha   = \sum\limits_q {\left[ {\hbar \omega _q \left| {\beta _{q\alpha } } \right|^2  - \frac{{i\hbar }}{2}\left( {\dot \beta _{q\alpha } \beta _{q\alpha }^ *   - \beta _{q\alpha } \dot \beta _{q\alpha }^ *  } \right)} \right]}\;. 
\end{equation}

\noindent Likewise, contracting Eq. (\ref{eq8}) with states $\left| {\alpha '} \right\rangle \left| {1_{q,\alpha '} } \right\rangle $ yields 

\[\left[ {i\hbar \dot \beta _{q'\alpha '}  - \hbar \omega _{q'} \beta _{q'\alpha '} } \right]\left\langle {{\alpha '}} \mathrel{\left | {\vphantom {{\alpha '} {\alpha '}}} \right. \kern-\nulldelimiterspace} {{\alpha '}} \right\rangle \delta _{0,\left\langle {{\alpha '}} \mathrel{\left | {\vphantom {{\alpha '} \Psi }}\right. \kern-\nulldelimiterspace}{\Psi } \right\rangle }  = \left\langle {\alpha '} \right.\left| {\left\langle {1_{q',\alpha '} } \right.\left| W \right|\left. {\beta _{\alpha '} } \right\rangle } \right|\left. {\alpha '} \right\rangle \delta _{0,\left\langle {{\alpha '}}\mathrel{\left | {\vphantom {{\alpha '} \Psi }}\right. \kern-\nulldelimiterspace}{\Psi } \right\rangle }  + \]

\begin{equation}
\label{eq20}
 +  \sum\limits_{\alpha  \ne \alpha '} {\left\{ {\left\langle {\alpha '} \right.\left| {H_{ex} } \right|\left. \alpha  \right\rangle \left\langle {{1_{q',\alpha '} }} \mathrel{\left | {\vphantom {{1_{q',\alpha '} } {\beta _\alpha  }}} \right. \kern-\nulldelimiterspace}{{\beta _\alpha  }} \right\rangle  + \left\langle {\alpha '} \right.\left| {\left\langle {1_{q',\alpha '} } \right.\left| W \right|\left. {\beta _\alpha  } \right\rangle } \right|\left. \alpha  \right\rangle  - i\hbar \left\langle {{\alpha '}}\mathrel{\left | {\vphantom {{\alpha '} {\dot \alpha }}}\right. \kern-\nulldelimiterspace}{{\dot \alpha }} \right\rangle \left\langle {{1_{q',\alpha '} }}\mathrel{\left | {\vphantom {{1_{q',\alpha '} } {\beta _\alpha  }}}\right. \kern-\nulldelimiterspace}{{\beta _\alpha  }} \right\rangle } \right\}}  \;.
\end{equation}

\noindent Using again ansatz (\ref{eq11}) for the interaction $W$ and Eq. (\ref{eq12}) in the form $
\left\langle {\alpha '} \right.\left| {G_{q,\alpha } } \right|\left. \alpha  \right\rangle  \sim \delta _{\alpha '\alpha } $, we note that

\[\sum\limits_{\alpha  \ne \alpha '} {\left\langle {\alpha '} \right.\left| {\left\langle {1_{q',\alpha '} } \right.\left| W \right|\left. {\beta _\alpha  } \right\rangle } \right|\left. \alpha  \right\rangle  = \sum\limits_{\alpha  \ne \alpha ',q} {\left[ {\left\langle {1_{q',\alpha '} } \right.\left| {b_q^{\dagger}  } \right|\left. {\beta _\alpha  } \right\rangle \left\langle {\alpha '} \right.\left| {G_{q,\alpha } } \right|\left. \alpha  \right\rangle  + \beta _{q\alpha } \left\langle {\alpha '} \right.\left| {w_q^{\dagger}  } \right|\left. \alpha  \right\rangle \left\langle {{1_{q',\alpha '} }}\mathrel{\left | {\vphantom {{1_{q',\alpha '} } {\beta _\alpha  }}} \right. \kern-\nulldelimiterspace}{{\beta _\alpha  }} \right\rangle } \right]} } \]

\begin{equation}
\label{eq21}
= \sum\limits_{\alpha  \ne \alpha ',q} {\beta _{q\alpha } \left\langle {\alpha '} \right.\left| {w_q^{\dagger}  } \right|\left. \alpha  \right\rangle \left\langle {{1_{q',\alpha '} }}\mathrel{\left | {\vphantom {{1_{q',\alpha '} } {\beta _\alpha  }}}\right. \kern-\nulldelimiterspace} {{\beta _\alpha  }} \right\rangle } = \sum\limits_{\alpha  \ne \alpha '} {\left\langle {\alpha '} \right.\left| {\left\langle {\beta _\alpha  } \right.\left| W \right|\left. {\beta _\alpha  } \right\rangle } \right|\left. \alpha  \right\rangle } \left\langle {{1_{q',\alpha '} }}
 \mathrel{\left | {\vphantom {{1_{q',\alpha '} } {\beta _\alpha  }}} \right. \kern \nulldelimiterspace}{{\beta _\alpha  }} \right\rangle \;.
\end{equation}

\noindent It becomes apparent now that the sum on the right-hand side of Eq. (\ref{eq20}) cancels by virtue of conditions (\ref{eq15}), while the remaining expression complements Eq. (\ref{eq12}), and shows that the exact form of Eqs.(\ref{eq13}) is

\begin{equation}
\label{eq22}
G_{q,\alpha } \left| \alpha  \right\rangle  = \left[ {i\hbar \dot \beta _{q\alpha }  - \hbar \omega _q \beta _{q\alpha } } \right]\left| \alpha  \right\rangle , \;\;\forall \alpha , q\;.
\end{equation}

Finally, it must be verified that the occupied lattice states remain orthogonal at all times [$\left\langle {{\alpha \left( t \right)}}\mathrel{\left | {\vphantom {{\alpha \left( t \right)} {\alpha '\left( t \right)}}}\right. \kern-\nulldelimiterspace}{{\alpha '\left( t \right)}} \right\rangle  \sim \delta _{\alpha ,\alpha '} $] under the propagation described by Eqs.(\ref{eq16}). Indeed, it is immediate that for any two occupied states, $\left\langle {\alpha } \mathrel{\left | {\vphantom {\alpha  \Psi }}\right. \kern-\nulldelimiterspace} {\Psi } \right\rangle  \ne 0$, $\left\langle {{\alpha '}} \mathrel{\left | {\vphantom {{\alpha '} \Psi }} \right. \kern-\nulldelimiterspace}{\Psi } \right\rangle  \ne 0$, 

\begin{equation}
\label{eq23}
i\hbar \left[ {\left\langle {{\alpha '}} \mathrel{\left | {\vphantom {{\alpha '} {\dot \alpha }}} \right. \kern-\nulldelimiterspace}{{\dot \alpha }} \right\rangle  + \left\langle {{\dot \alpha '}}\mathrel{\left | {\vphantom {{\dot \alpha '} \alpha }}\right. \kern-\nulldelimiterspace}{\alpha } \right\rangle } \right] = \left\langle {\alpha '} \right|\left\langle {\beta _\alpha  } \right.\left| W \right|\left. {\beta _\alpha  } \right\rangle \left| \alpha  \right\rangle  - \left\langle {\alpha '} \right|\left\langle {\beta _{\alpha '} } \right.\left| W \right|\left. {\beta _{\alpha '} } \right\rangle \left| \alpha  \right\rangle \;,
\end{equation}

\noindent which, in view of Eqs.(\ref{eq17}) , (\ref{eq22})  and (\ref{eq14}), becomes succesively

\[ i\hbar \frac{d}{{dt}}\left\langle {{\alpha '}}\mathrel{\left | {\vphantom {{\alpha '} \alpha }}\right. \kern-\nulldelimiterspace}{\alpha } \right\rangle  = \sum\limits_q {\left[ {\beta _{q\alpha } \left\langle {\alpha '} \right. \left|  {w_q^{\dagger}  }  \right| \left. \alpha  \right\rangle  - \beta _{q\alpha '}^*  \left\langle {\alpha '} \right. \left| {w_q } \right| \left. \alpha  \right\rangle } \right]}   \]

\begin{equation}
\label{eq24}
=  \sum\limits_{\rm q} {\left[ {\beta _{q\alpha } \beta _{q'\alpha '}^*  \left\langle {\alpha '} \right.\left| {u_{qq'}^{\dagger}  } \right|\left. \alpha  \right\rangle  - \beta _{q\alpha '}^*  \beta _{q'\alpha } \left\langle {\alpha '} \right.\left| {u_{qq'} } \right|\left. \alpha  \right\rangle } \right]}  = \sum\limits_{\rm q} {\beta _{q\alpha '}^*  \beta _{q'\alpha } \left[ {\left\langle {\alpha '} \right.\left| {u_{q'q}^{\dagger}  } \right|\left. \alpha  \right\rangle  - \left\langle {\alpha '} \right.\left| {u_{qq'} } \right|\left. \alpha  \right\rangle } \right]}  = 0 \;.
\end{equation}

\noindent The last line above follows from the condition that $u_{q'q}^{\dagger}   = u_{qq'} $ in order to secure the hermiticity of $W$. \\

To sum our results to this point, a  (generalized) $D_{1}$ state $\left| \Psi  \right\rangle $
 [Eq. (\ref{eq4})] describes a self-consistent lattice-phonon dynamics if and only if

a) the interaction $W$ is of the form (\ref{eq11});

b) the lattice states $\left| \alpha  \right\rangle $ are driven by the effective Schroedinger equations (\ref{eq16}) and each satisfy, simulteneously, a (large) number of DFS-like constraints imposed by the eigenvalue equations (\ref{eq22}) for the operators $G_{q,\alpha}$;

c) the phonon displacement parameters $\beta_{q\alpha}$ obey the evolution equations 

\begin{equation}
\label{eq25}
i\hbar \dot \beta _{q\alpha }  - \hbar \omega _q \beta _{q\alpha }  = \gamma _{q,\alpha } \left( {\beta _\alpha  ,w,u} \right), \;\;\forall\; q,\alpha \;,
\end{equation}

\noindent where $\gamma_{q,\alpha}$ denotes an eigenvalue of $G_{q,\alpha}$ (assuming any exists) and we have indicated explicitly the dependence on the set of $\beta_{q\alpha}$'s and on the interaction factors $w_{q}$ and $u_{qq'}$. 

As suggested earlier, self-consistency is seen to require that the propagation of individual $\left| \alpha  \right\rangle $'s and their associated phonon parameters $\left\{ {\beta _{q\alpha } } \right\}$ be decoupled from similar terms. This rather strong result shows that, under quite general conditions, a lattice-phonon system can support $D_{1}$ states if and only if it can support the component orthogonal $D_{2}$ states $\left| {\Psi _\alpha  } \right\rangle  = \left| \alpha  \right\rangle \left| {\beta _\alpha  } \right\rangle $ individually. In particular, it ensues that a given system can sustain standard $D_{1}$ states

\begin{equation}
\label{eq26}
\left| {\Psi _1 } \right\rangle  = \sum\limits_{n = 1}^N {\varphi _n c_n^{\dagger}  \left| {\beta _n } \right\rangle \left| 0 \right\rangle _{ex} } 
\end{equation}

\noindent if and only if it can sustain on-site, single-quantum coherent states $\left| {\Phi _n } \right\rangle  = c_n^{\dagger}  \left| 0 \right\rangle _{ex} \left| {\beta _n } \right\rangle $. Even under such circumstances, it turns out that the $D_{1}$ states can only yield static lattice distributions [static 'solitons']. Indeed, according to Eq. (\ref{eq24}), $\left\langle {\alpha }\mathrel{\left | {\vphantom {\alpha  \alpha }}\right. \kern-\nulldelimiterspace}{\alpha } \right\rangle  = const$ [consistent with the unitary evolution of $\left| \alpha  \right\rangle $] and, correspondingly, the probabilities $\left| {\varphi _n } \right|^2 $ in any self-consistent, standard $D_{1}$  state [Eq. (\ref{eq26})]  are conserved in time. The same holds true for higher order (multiquanta) $D_{1}$ states based on on-site $\left| \alpha  \right\rangle $ states, as well as for their superpositions. As a corollary, mobile lattice distributions can arise if and only if the driving Hamiltonian accomodates $D_{2}$ states with delocalized lattice components.

Let us now note that, while Eqs.(\ref{eq16}) are just the usual Davydov equations for the [$D_{2}$] lattice factor states, Eqs.(\ref{eq25}) do not involve the $\left| \alpha  \right\rangle $'s themselves, hence no parameters defining these states can appear in the evolution equations for the $\beta_{q\alpha}$'s. In other words, a solution for the  $\beta_{q\alpha}$'s, when it exists, will not show a functional dependence on the parameters of $\left| \alpha  \right\rangle $, and substitution of the explicit expressions for the $\beta_{q\alpha}$'s into the Hamiltonian of Eq. (\ref{eq19}) cannot lead to nonlinear equations for $\left| \alpha  \right\rangle $. 

To clarify the relation between this result and Davydov's soliton equations, rewrite the equations of motion for the $\beta _{q\alpha } $'s as [Eq. (\ref{eq22}), $\left\langle {\alpha }\mathrel{\left | {\vphantom {\alpha  \alpha }}\right. \kern-\nulldelimiterspace} {\alpha } \right\rangle  = 1$]

\begin{equation}
\label{eq27}
i\hbar \dot \beta _{q\alpha }  - \hbar \omega _q \beta _{q\alpha }  = \left\langle \alpha  \right|G_{q,\alpha } \left| \alpha  \right\rangle {\rm . }
\end{equation}

\noindent A linear parametrization of $\left| \alpha  \right\rangle $ in terms of some fixed lattice basis states, $\left| \alpha  \right\rangle  = \sum\limits_k {\alpha _k \left| k \right\rangle } $, will induce, apparently, nonlinear equations typical of Davydov's theory. In fact, it can be verified that Eqs.(\ref{eq16}) and (\ref{eq27}) do recover the $D_{2}$ form of the soliton equations for the appropriate Hamiltonian and choice of $\left| \alpha  \right\rangle $.  However, Eqs.(\ref{eq16}) and (\ref{eq27}) cannot describe a self-consistent dynamics unless complemented by the requirement that $\left| \alpha  \right\rangle $ also be an eigenfunction of $G_{q,\alpha } $. But, in that case, the average on the right-hand side of Eq. (\ref{eq27}) reduces to the corresponding eigenvalue, and the propagation of $\left| \alpha  \right\rangle $ becomes linear. It is interesting that although the exact form of Eqs. (\ref{eq22}) is known in the literature  for certain extensions of Davydov's model [see, e.g., Ref. \cite{forner}], their interpretation was merely as a particular type of nonlinear equations for the phonon parameters. Of course, the linear effective evolution of the lattice states $\left| \alpha  \right\rangle $ does not preclude, by itself, nontrivial phenomena in the lattice dynamics. Provided appropriate solutions exist, it is conceivable that nonlinear effects can yet arise, and be modulated by the phonon contribution. However, the circumstances leading to such effects are, obviously, rather restrictive.

The constraint that $\left| \alpha  \right\rangle $ be a (simultaneous) eigenfunction of the operators $G_{q\alpha } $ can be understood also as a restriction on the allowed initial state $\left| \alpha  \right\rangle $, as seen by substituting $\left| {\alpha \left( t \right)} \right\rangle  = U_\alpha  \left( t \right)\left| {\alpha \left( 0 \right)} \right\rangle $, with $U_\alpha  \left( t \right)$ the unitary evolution operator corresponding to the effective Hamiltonian of Eq. (\ref{eq16}). Further, for given $\alpha $, the set of constraints (\ref{eq22}) is equivalent to the single initial-state constraint

\begin{equation}
\label{eq28}
\Lambda _\alpha  \left| {\alpha \left( 0 \right)} \right\rangle  = 0 \;,
\end{equation}

\noindent for 

\begin{equation}
\label{eq29}
\Lambda _\alpha   = \int\limits_0^\infty  {dt\sum\limits_q {\left[ {\bar G_{q,\alpha }^{\dagger}  \left( t \right) - \gamma _{q,\alpha }^ *  \left( t \right)} \right]\left[ {\bar G_{q,\alpha } \left( t \right) - \gamma _{q,\alpha } \left( t \right)} \right]} } \;,
\end{equation}

\noindent with

\begin{equation}
\label{eq30}
\bar G_{q,\alpha } \left( t \right) = U_\alpha ^{\dagger}  \left( t \right)G_{q,\alpha } \left( t \right)U_\alpha  \left( t \right)\;.
\end{equation}

\noindent In this point of view, the existence of self-consistent $D_{1}$ states is conditioned by the existence of an appropriate subspace of acceptable initial lattice states, determined by Eqs.(\ref{eq28}) -(\ref{eq30}). When such a subspace exists, the propagation of any $D_{1}$ state [Eq. (\ref{eq4})] with initial lattice states belonging to this subspace {\it may} be described by the generalized Davydov equations (\ref{eq16}) and (\ref{eq27}), but, as noted above, the apparent nonlinearity of the underlying lattice dynamics cancels identically. Equations (\ref{eq27}) reduce in fact to Eqs.(\ref{eq25}). In Sec. IV it will be shown that self-consistent propagation at finite-temperature further restricts the allowable forms of lattice-phonon interactions, and sets final bounds on the subspace of permitted initial states.

\section{Density matrix formal framework}
\label{Sec3}
 
The approach of the previous section will be extended to the finite-temperature domain in a framework that borrows heavily from the thermofield dynamics (TFD) construct \cite{thermo} [a somewhat related formalism is detailed in \cite{lawrie}]. Specifically, the total density-matrix for the lattice-phonon system is written as $\hat \rho  = \hat \gamma \hat \gamma ^{\dagger}  $, where the nonhermitian state operator $\hat \gamma $ is defined up to a unitary gauge transformation of the kind $\hat \gamma  \to \hat \gamma U\left( t \right)$, $UU^{\dagger}   = U^{\dagger}  U = I$. In other words, any two distinct state operators $\hat \gamma ,{\rm  }\hat \gamma '$ corresponding to the same density matrix are necessarily related by $\hat \gamma ' = \hat \gamma U\left( t \right)$. From the evolved density-matrix, written in the form

\[
\hat \rho \left( t \right) = \exp \left[ { - \frac{i}{\hbar }H{\rm  }t} \right]\gamma \left( 0 \right)U\left( t \right) \cdot U^{\dagger}  \left( t \right)\gamma ^{\dagger}  \left( 0 \right)\exp \left[ {\frac{i}{\hbar }H{\rm  }t} \right]\;,
\]

\noindent one can infer that 

\[
\gamma \left( t \right) = \exp \left[ { - \frac{i}{\hbar }H{\rm  }t} \right]\gamma \left( 0 \right)U\left( t \right)  
\]

\noindent satisfies the von Neumann-like equation 

\begin{equation}
\label{eq31}
i\hbar \frac{{d\hat \gamma }}{{dt}} = H \cdot \hat \gamma  - \hat \gamma  \cdot K\;\;,
\end{equation}

\noindent with $K$ the [arbitrary] hermitian generator of the unitary gauge factor $U\left( t \right)$. In superoperator notation, Eq. (\ref{eq31}) reads

\begin{equation}
\label{eq32}
i\hbar \frac{d\hat \gamma }{dt} = \left( {\bf H} - {\bf \tilde K} \right)\hat \gamma \;,
\end{equation}

\noindent where ${\bf H}$ is the superoperator defined by the Hamiltonian $H$, ${\bf H}\hat \gamma  = H \cdot \hat \gamma $, and ${\bf \tilde K}$ is given by ${\bf \tilde K}\hat \gamma  = \hat \gamma  \cdot K$, $\left[ {{\bf H},{\rm  }{\bf \tilde K}} \right] = 0$. As in TFD, the tilde notation is used here for the tilde conjugate of a superoperator ${\bf A}$, linear or not, introduced as \cite{thermo}

\begin{equation}
\label{eq33}
\left[ {{\bf A}\left( {\hat \alpha } \right)} \right]^{\dagger}   = {\bf \tilde A}\left( {\hat \alpha ^{\dagger}  } \right)\;.
\end{equation}

\noindent If ${\bf A}\left( {\hat \alpha } \right) = A \cdot \hat \alpha $ for some linear operator $A$, Eq. (\ref{eq33}) yields ${\bf \tilde A}\left( {\hat \alpha } \right) = \hat \alpha  \cdot A^{\dagger}  $, which reduces to ${\bf \tilde A}\left( {\hat \alpha } \right) = \hat \alpha  \cdot A$ when $A$ is self-adjoint. It is immediate that the tilde operation is distributive against the usual addition and multiplication of linear operators, but antilinear with respect to the multiplication by scalars. Also, tilde-symmetric superoperators map self-adjoint operators onto self-adjoint operators.

Obviously, from Eq. (\ref{eq31}) one can retrieve the von Neumann equation for the density-matrix, which in superoperator form reads $i\hbar d\hat \rho /dt = \left( {{\bf H} - {\bf \tilde H}} \right)\hat \rho $. The gauge generator $K$ does not bear any physical significance, but, as will be seen in the next section, may prove instrumental in streamlining the calculation, although it can suffer restrictions under a given ansatz for $\hat \gamma $. Note that while ${\bf H}$ can contain only direct superoperators generated by the usual observables [or creation and annihilation operators], ${\bf \tilde K}$ can only contain the corresponding tilde conjugates.

If the space of linear operators is endowed with the Hilbert space structure induced by the inner product 

\begin{equation}
\label{eq34}
\left( {\hat \sigma |\hat \omega } \right) = Tr\left( {\hat \sigma ^{\dagger}  \hat \omega } \right)\;,
\end{equation}

\noindent $\hat \gamma $ spans the sphere of unit norm operators [$\left( {\hat \gamma |\hat \gamma } \right) = 1$] , and the average of an observable  becomes 

\begin{equation}
\label{eq35}
Tr\left( {O\hat \rho } \right) = Tr\left( {O\hat \gamma \hat \gamma ^{\dagger}  } \right) = \left( {\hat \gamma } \right.|{\bf O}|\left. {\hat \gamma } \right)\;.
\end{equation}

\noindent An orthonormal operator basis $\left\{ {\hat \sigma } \right\}$, $\left( {\hat \sigma |\hat \sigma '} \right) = \delta _{\sigma \sigma '} $, determines a basis representation of $\hat \gamma $ as

\begin{equation}
\label{eq36}
\hat \gamma  = \sum\limits_\sigma  {\left( {\hat \sigma |\hat \gamma } \right){\rm  }\hat \sigma } \;,
\end{equation}

\noindent and similarly for observables [superoperators]. Special attention receives the operator equivalent of the Hilbert space Fock basis, which implements the operator Fock space. If $a$ and $a^{\dagger}$ are boson annihilation and creation operators for the Hilbert space vacuum $\left| 0 \right\rangle $, the annihilation and creation [superoperator] counterparts for the operator vacuum [projector] $\left| 0 \right\rangle \left\langle 0 \right|$ are ${\bf a},{\rm  }{\bf \tilde a}$ and, respectively, ${\bf a}^{\dagger}  ,{\rm  }{\bf \tilde a}^{\dagger}  $. The corresponding operator Fock basis follows simply as the exterior product of vectors of the Hilbert space Fock basis,  

\begin{equation}
\label{eq37}
|\left. {n,\tilde m} \right) = \frac{1}{{\sqrt {n!m!} }}\left( {{\bf a}^{\dagger}  } \right)^n \left( {{\bf \tilde a}^{\dagger}  } \right)^m \left| 0 \right\rangle \left\langle 0 \right| \equiv \frac{1}{{\sqrt {n!m!} }}\left( {a^{\dagger}  } \right)^n \left| 0 \right\rangle \left\langle 0 \right|a^m  = \left| n \right\rangle \left\langle m \right| \;.
\end{equation}

\noindent Applying now the central idea of TFD, note that a boson thermal state operator

\[
\hat \gamma _T  = Z^{-1/2} \exp \left[ { - (\hbar \omega /2k_B T)a^{\dagger}  a} \right]
\]

\noindent is related to the zero-temperature vacuum by the unitary, tilde-symmetric (super)transformation \cite{thermo}

\begin{equation}
\label{eq39}
\hat \gamma _T  = {\bf U}\left( T \right)\left| 0 \right\rangle \left\langle 0 \right| \;,
\end{equation}

\[
{\bf U}\left( T \right) = \exp \left[ {\left( {{\bf a}^{\dagger}  {\bf \tilde a}^{\dagger}   - {\bf a\tilde a}} \right)\theta } \right]       \;,
\]

\noindent with $\tanh \theta  = \exp \left[ { - \hbar \omega /2k_B T} \right]$. Hence, it is itself vacuum for the thermal annihilation (super)operators

\begin{mathletters}
\label{eq40}

\begin{equation}
{\bf A} \equiv {\bf U}\left( T \right){\rm  }{\bf a}{\rm  }{\bf U}^{\dagger}  \left( T \right) = {\bf a}\cosh \theta  - {\bf \tilde a}^{\dagger}  \sinh \theta \;,
\end{equation}

\begin{equation}
{\bf \tilde A} \equiv {\bf U}\left( T \right){\rm  }{\bf \tilde a}{\rm  }{\bf U}^{\dagger}  \left( T \right) =  - {\bf a}^{\dagger}  \sinh \theta  + {\bf \tilde a}\cosh \theta \;.
\end{equation}

\end{mathletters}

\noindent Just as in the zero-temperature case, the corresponding operator Fock space can be constructed on the 'thermal vacuum' $\hat \gamma _T $ by using the thermal creation (super)operators ${\bf A}^{\dagger}$ and ${\bf \tilde A}^{\dagger}$.

The advantage of using the state operator $\hat \gamma $ and Eq. (2.2) over the density-matrix and the von Neuman equation comes from the evident analogy between this framework and the usual Hilbert space formalism. In addition, for the problem at hand, our Gaussian density-matrix ansatz (see below) has a simpler, if somewhat abstract, expression for $\hat \gamma $.

\section{Self-consistent density-matrix coherent-product [$D_1$] states}
\label{Sec4}

The zero-temperature $D_1 $ ansatz for pure wave functions (state vectors) is generalized here to the product ansatz 

\begin{equation}
\label{eq42}
\hat \gamma  = \sum\limits_\alpha  {\hat \alpha  \cdot \hat \gamma _\alpha   \cdot U} 
\end{equation}

\noindent for the state operator. In direct analogy to the pure state ansatz (\ref{eq4}), the lattice state operators $\hat \alpha $ are assumed orthogonal, but not normalized, in the sense of the operator scalar product $\left( {\hat \alpha } \right.|\left. {\hat \alpha '} \right) = Tr\left( {\hat \alpha ^{\dagger}  \hat \alpha '} \right) \sim \delta _{\alpha '\alpha } $, while the phonon state operators $\hat \gamma _\alpha  $ are given the thermal Gaussian [coherent] form 

\begin{equation}
\label{eq43}
\hat \gamma _\alpha   = \prod\limits_q {\hat \gamma _{q,\alpha } }  = \prod\limits_q {\rm  } \exp \left[ {\beta _{q\alpha } b_q^{\dagger}   - \beta _{q\alpha }^* b_q } \right]\hat \gamma _{q,T} \;.
\end{equation}

\noindent with

\begin{equation}
\label{eq44}
\hat \gamma _{q,T}  = \frac{1}{{\sqrt {Z_q } }}\exp \left[ { - \frac{{\hbar \omega _q }}{{2k_B T}}b_q^{\dagger}  b_q } \right]\;.
\end{equation}

\noindent The overall conservation of probability requires $\sum\limits_\alpha  {\left( {\hat \alpha } \right.|\left. {\hat \alpha } \right)}  = 1$. Also as in the pure state case, the $\hat \alpha $'s will be considered a subset of an orthogonal lattice operator basis $\left\{ {\hat \alpha } \right\}$ and the noncontributing $\hat \alpha $'s, $\left( {\hat \alpha } \right.|\left. {\hat \gamma } \right) = 0$, will be assigned by default null phonon displacements, $\beta _{q\alpha }  = 0$. 

The ansatz (\ref{eq42}) corresponds to the density-matrix

\begin{equation}
\label{eq45}
\hat \rho   = \sum\limits_{\alpha ',\alpha } {\left( {\hat \alpha '\hat \alpha ^{\dagger}  } \right)} \left( {\hat \gamma _{\alpha '} \hat \gamma _\alpha ^{\dagger}  } \right) \;,
\end{equation}

\noindent which can be understood, in general, as a strongly entangled [incoherent] superposition of phonon Gaussian states. Indeed, averaging over the lattice degrees of freedom yields the phonon density-matrix as

\begin{equation}
\label{eq46}
\hat \rho _{ph}  = \sum\limits_\alpha  {\left( {\hat \alpha |\hat \alpha } \right)\prod\limits_q {\frac{1}{{Z_q }}\exp \left[ { - \frac{{\hbar \omega _q }}{{k_B T}}\left( {b_q^{\dagger}   - \beta _{q\alpha }^* } \right)\left( {b_q  - \beta _{q\alpha } } \;.\right)} \right]} } 
\end{equation}

\noindent When the lattice factors $\hat \alpha $ correspond to mutually orthogonal pure states, i.e., $\hat \alpha  = \left| \alpha  \right\rangle \left\langle \alpha  \right|U$ for some common unitary factor $U$, $\left\langle {\alpha } \mathrel{\left | {\vphantom {\alpha  {\alpha '}}} \right. \kern-\nulldelimiterspace} {{\alpha '}} \right\rangle  \sim \delta _{\alpha \alpha '} $, one has $\hat \alpha '\hat \alpha ^{\dagger}   = \delta _{\alpha '\alpha } \left| \alpha  \right\rangle \left\langle \alpha  \right|$ and Eq. (\ref{eq45}) acquires the simpler and more transparent form

\begin{equation}
\label{eq47}
\hat \rho  = \sum\limits_\alpha  {\left| \alpha  \right\rangle } \prod\limits_q {\frac{1}{{Z_q }}\exp \left[ {\beta _{q\alpha } b_q^{\dagger}   - \beta _{q\alpha }^* b_q } \right]\exp \left[ { - \frac{{\hbar \omega _q }}{{k_B T}}b_q^{\dagger}  b_q } \right]} \exp \left[ { - \beta _{q\alpha } b_q^{\dagger}   + \beta _{q\alpha }^* b_q } \right]\left\langle \alpha  \right|\;.
\end{equation}

\noindent If the [unnormalized] states $\left| \alpha  \right\rangle $ are set proportional to the on-site, one-quantum states $\left| {1_n } \right\rangle  \equiv c_n^{\dagger}  \left| 0 \right\rangle $, expression (\ref{eq47}) recovers Davydov's original finite-temperature ansatz \cite{Davydov}. At zero temperature the phonon contribution reduces to $\hat \gamma _\alpha   \to \left| {\beta _\alpha  } \right\rangle \left\langle 0 \right|_{ph} $ and the ansatz (\ref{eq42}) yields

\[
\hat \gamma  = \sum\limits_\alpha  {\hat \alpha  \cdot \left| {\beta _\alpha  } \right\rangle \left\langle 0 \right|_{ph} U}\;. 
\]

\noindent For lattice factors $\hat \alpha  = \left| \alpha  \right\rangle \left\langle \chi  \right|$, with a common $\left\langle \chi  \right|$, $\left\langle {\chi } \mathrel{\left | {\vphantom {\chi  \chi }} \right. \kern-\nulldelimiterspace} {\chi } \right\rangle  = 1$, the state operator becomes $\hat \gamma  = \left[ {\sum\limits_\alpha  {\left| \alpha  \right\rangle \left| {\beta _\alpha  } \right\rangle } } \right]\left\langle \chi  \right|\left\langle 0 \right|_{ph} $
 and corresponds to a pure D1 state of the type (\ref{eq4}). Even more remarkably, when all displacement parameters vanish, $\beta _{q\alpha }  = 0$, all $\hat \gamma _\alpha  $ reduce to the thermal state operator $\hat \gamma _T  = \prod\limits_q {\hat \gamma _{q,T} } $, and the density-matrix (\ref{eq45}) acquires the familiar product form 

\begin{equation}
\label{eq48}
\hat \rho  = \hat \rho _{lat}  \cdot \hat \rho _{ph,T} \;,
\end{equation}

\noindent where $\hat{ \rho} _{lat}  = \left[ \sum\limits_\alpha  {\hat \alpha }  \right] \cdot \left[ \sum\limits_\alpha  {\hat \alpha }\right]^{\dagger}$ and $\hat{\rho} _{ph,T}  = \prod\limits_q {(Z_q)^{-1}\exp \left[  - (\hbar \omega _q /k_B T)b_q^{\dagger}  b_q  \right] } $. Since the lattice state operator $\left[ \sum\limits_\alpha {\hat \alpha } \right]$ is quite arbitrary, the ansatz (\ref{eq42}) is seen to cover all situations where a product state as in Eq. (\ref{eq48}) evolves into an entangled mixture described by Eq. (\ref{eq45}). Note that the ansatz (\ref{eq42}) accounts for a much wider class of evolutions, because, in general, the displacement parameters $\beta _{q \alpha}$ need not vanish simultaneously at any time. \\

Let us now introduce, for each phonon mode $q$, the thermal annihilation (super)operators

\begin{mathletters}
\label{eq49}

\begin{equation}
{\bf B}_q  = {\bf b}_q \cosh \theta _q  - {\bf \tilde b}_q^{\dagger}  \sinh \theta _q \;,
\end{equation}

\begin{equation}
{\bf \tilde B}_q  =  - {\bf b}_q^{\dagger}  \sinh \theta _q  + {\bf \tilde b}_q \cosh \theta _q \;,
\end{equation}

\end{mathletters}

\noindent with $\tanh \theta _q  = \exp \left[ { - \hbar \omega _q /2k_B T} \right]$. Upon substituting the reciprocal transformations into $\hat \gamma _{q,\alpha }  = \exp \left[ {\beta _{q\alpha } {\bf b}_q^{\dagger}   - \beta _{q\alpha }^* {\bf b}_q } \right]\hat \gamma _{q,T}$, it is seen that $\hat \gamma _\alpha   = \prod\limits_q {\hat \gamma _{q,\alpha } } $ represents a displaced vacuum for ${\bf B}_q $ and ${\bf \tilde B}_q $, since

\begin{mathletters}
\label{eq52}

\begin{equation}
{\bf B}_q \hat \gamma _\alpha   = \beta _{q\alpha } \cosh \theta _q {\rm  }\hat \gamma _\alpha \;,
\end{equation}

\begin{equation}
{\bf \tilde B}_q \hat \gamma _\alpha   =  - \beta _{q\alpha } \sinh \theta _q {\rm  }\hat \gamma _\alpha \;.
\end{equation}

\end{mathletters}

\noindent The displaced, orthonormal Fock basis constructed for each $\hat \gamma _\alpha  $  will be denoted $|\left. { \ldots n_{q,\alpha } ,\tilde m_{q,\alpha }  \ldots } \right)$ and reads explicitly 

\begin{equation}
\label{eq53}
|\left. { \ldots n_{q,\alpha } ,\tilde m_{q,\alpha }  \ldots } \right) = \prod\limits_q {\frac{1}{\sqrt {n_q !}}\frac{1}{\sqrt {m_q !}} \left( {{\bf B}_q^{\dagger}   - \beta _{q\alpha }^* \cosh \theta } \right)^{n_q } \left( {{\bf \tilde B}_q^{\dagger}   + \beta _{q\alpha }^* \sinh \theta } \right)^{m_q } \hat \gamma _\alpha  }\;.
\end{equation}

\noindent As in the pure state framework, the set $\left\{ {\hat \alpha  \cdot |\left. { \ldots n_{q,\alpha } ,\tilde m_{q,\alpha }  \ldots } \right)} \right\}$ provides an orthogonal operator basis for the total lattice-phonon system, and the analogy can be carried further, step by step. The self-consistency conditions for the ansatz (\ref{eq42}) can be derived now from the expansion of the associated equation of motion (\ref{eq32}) in the basis $\left\{ {\hat \alpha  \cdot |\left. { \ldots n_{q,\alpha } ,\tilde m_{q,\alpha }  \ldots } \right)} \right\}$.

But, a word is in order first regarding the gauge generator ${\bf \tilde K}$. One should note that substitution of the reciprocals of transformations (\ref{eq49}) in the bare phonon Hamiltonian ${\bf H}_{ph}  = \sum\limits_q {\hbar \omega _q {\bf b}_q^{\dagger}  {\bf b}_q } $ produces nondiagonal products of ${\bf B}$'s and ${\bf \tilde B}$'s , which may prove cumbersome. But if it is observed, as in TFD \cite{thermo}, that 

\[
{\bf b}_q^{\dagger}  {\bf b}_q  - {\bf \tilde b}_q^{\dagger}  {\bf \tilde b}_q  = {\bf B}_q^{\dagger}  {\bf B}_q  - {\bf \tilde B}_q^{\dagger}  {\bf \tilde B}_q \;,
\]

\noindent the gauge generator can be suitably rewritten 

\begin{equation}
\label{eq54}
{\bf \tilde K} \to {\bf \tilde K} - \sum\limits_q {\hbar \omega _q {\bf \tilde b}_q^{\dagger}  {\bf \tilde b}_q } \;,
\end{equation}

\noindent which brings the free phonon Hamiltonian for Eq. (\ref{eq32}) to the invariant form 

\begin{equation}
\label{eq55}
{\bf H}_{ph}  - {\bf \tilde H}_{ph}  = \sum\limits_q {\hbar \omega _q \left( {{\bf b}_q^{\dagger}  {\bf b}_q  - {\bf \tilde b}_q^{\dagger}  {\bf \tilde b}_q } \right)}\;.
\end{equation}

\noindent The equation of motion (\ref{eq32}) acquires thus the particular form

\begin{equation}
\label{eq56}
i\hbar \frac{{d\hat \gamma }}{{dt}} = \left[ {{\bf H}_{lat}  + {\bf W} + \left( {{\bf H}_{ph}  - {\bf \tilde H}_{ph} } \right) - {\bf \tilde K}} \right]\hat \gamma \;.
\end{equation}

Subsequent substitution of the product ansatz (\ref{eq42}) produces

\[
i\hbar \sum\limits_a {\frac{{d\hat \alpha }}{{dt}}\hat \gamma _\alpha  }  + \frac{{i\hbar }}{2}\sum\limits_{\alpha ,q} {\left[ {\dot \beta _{q\alpha } \beta _{q\alpha }^*  - \beta _{q\alpha } \dot \beta _{q\alpha }^* } \right]{\rm  }} \hat \alpha  \cdot \hat \gamma _\alpha   +  
\]

\[
  + i\hbar {\rm  }\sum\limits_{\alpha {\rm ,q}} {\dot \beta _{{\rm q}\alpha } \cosh \theta _q {\rm  }\hat \alpha  \cdot |\left. {1_{q,\alpha } } \right)}  - i\hbar {\rm  }\sum\limits_{\alpha {\rm ,q}} {\dot \beta _{q\alpha }^* \sinh \theta _q {\rm  }\hat \alpha  \cdot |\left. {\tilde 1_{q,\alpha } } \right)}  =  
\]

\[
  = \sum\limits_\alpha  {\left( {{\bf H}_{lat} \hat \alpha } \right) \cdot \hat \gamma _\alpha  }  + \sum\limits_\alpha  {\left( {{\bf W} - {\bf \tilde K}} \right)\hat \alpha  \cdot \hat \gamma _\alpha  }  + \sum\limits_{\alpha ,q} {\hbar \omega _q \left| {\beta _{q\alpha } } \right|^2 \hat \alpha  \cdot \hat \gamma _\alpha  }  + 
\]

\begin{equation}
\label{eq57}
  + {\rm  }\sum\limits_{\alpha ,q} {\hbar \omega _q \beta _{q\alpha } \cosh \theta _q {\rm  }\hat \alpha  \cdot |\left. {1_{q,\alpha } } \right)}  + {\rm  }\sum\limits_{\alpha ,q} {\hbar \omega _q \beta _{q\alpha }^* \sinh \theta _q {\rm  }\hat \alpha  \cdot |\left. {\tilde 1_{q,\alpha } } \right)} \;, 
\end{equation}

\noindent which is to be contracted, succesively, with each of the basis operators $\left\{ {\hat \alpha  \cdot |\left. { \ldots n_{q,\alpha } ,\tilde m_{q,\alpha }  \ldots } \right)} \right\}$. \\

The results of the preceding section can be extended straightforwardly, mostly by obvious substitutions. For instance, the contraction with operators $\left\{ {\hat \alpha ' \cdot |\left. { \ldots n_{q,\alpha '} ,\tilde m_{q,\alpha '}  \ldots } \right)} \right\}$ carrying more than one excited 'quantum', i.e., $\sum\limits_q {\left( {n_q  + \tilde m_q } \right)}  > 1$, shows that the interaction-gauge term $\left( {{\bf W} - {\bf \tilde K}} \right)$ must be such that

\begin{equation}
\label{eq58}
\left( { \ldots n_{q,\alpha } ,\tilde m_{q,\alpha }  \ldots } \right.|{\bf W} - {\bf \tilde K}|\left. {\gamma _\alpha  } \right) = 0\;.
\end{equation}

\noindent Hence the allowed phonon dependencies are zero order, first order or bilinear of the type  ${\bf B}_q^{\dagger}  {\bf B}_{q'} $, ${\bf \tilde B}_q^{\dagger}  {\bf B}_{q'} $ and the tilde conjugates. However, it can be checked that, due to the specific character of ${\bf W}$ and ${\bf \tilde K}$ [as functionals of right-acting and left-acting operators], the bilinear terms can only arise accompanied by prohibited terms of the form $\left( {\cosh \theta _q \sinh \theta _{q'} {\bf B}_q^{\dagger}  {\bf \tilde B}_{q'}^{\dagger}  } \right)$ and $\left( {\cosh \theta _q \sinh \theta _{q'} {\bf B}_q {\bf \tilde B}_{q'} } \right)$. For this reason, the allowed expressions for ${\bf W}$ and ${\bf \tilde K}$ at finite-temperatures [$\sinh \theta _q  \ne 0$] reduce to the very simple forms [in terms of zero temperature operators]

\begin{equation}
\label{eq59}
{\bf W} = \sum\limits_q {\left( {{\bf w}_q {\bf b}_q^{\dagger}   + {\bf w}_q^{\dagger}  {\bf b}_q } \right)} 
\end{equation}

\noindent and, respectively,

\begin{equation}
\label{eq60}
{\bf \tilde K} = {\bf \tilde K}_{lat}  + \sum\limits_q {\left( {{\bf \tilde v}_q {\bf \tilde b}_q^{\dagger}   + {\bf \tilde v}_q^{\dagger}  {\bf \tilde b}_q } \right)} \;,
\end{equation}

\noindent where the zero-order term in ${\bf W}$ is assumed subsumed in ${\bf H}_{lat} $, and ${\bf \tilde K}_{lat} $, ${\bf w}_q $ and ${\bf v}_q $ are lattice operators, functionals of $c_n $ and $c_n^{\dagger}  $ only. At zero temperature, bilinear coupling terms can still coexist and the outcome parallels the results of Sec. II. Let us consider here the finite-temperature case. 

From the same contraction procedure it also follows that 

\begin{equation}
\label{eq61}
\left( {\hat \alpha '} \right.|\left( {1_{q,\alpha } } \right.|{\bf W} - {\bf \tilde K}|\left. {\hat \gamma _\alpha  } \right)|\left. {\hat \alpha } \right) = \left( {\hat \alpha '} \right.|\left( {\tilde 1_{q,\alpha } } \right.|{\bf W} - {\bf \tilde K}|\left. {\hat \gamma _\alpha  } \right)|\left. {\hat \alpha } \right) = 0 , \;\forall q, \;\alpha  \ne \alpha '
\end{equation}

\noindent and

\begin{equation}
\label{eq62}
i\hbar \left( { \hat \alpha '\left|{ \frac{d\hat \alpha}{dt} }\right. } \right) = \left( {\hat \alpha '} \right.|{\bf H}_{lat} |\left. {\hat \alpha } \right) + \left( {\hat \alpha '} \right.|\left( {\hat \gamma _\alpha  } \right.|{\bf W} - {\bf \tilde K}|\left. {\hat \gamma _\alpha  } \right)|\left. {\hat \alpha } \right),  \;\forall \;\alpha  \ne \alpha ', \;\left( {\hat \alpha } \right.|\left. {\hat \gamma } \right) \ne 0 \;.
\end{equation}

\noindent Using expressions (\ref{eq59}) and (\ref{eq60}), one obtains from Eqs.(\ref{eq61})

\begin{mathletters}
\label{eq63}

\begin{equation}
\left( {\hat \alpha '} \right.|{\bf w}_q |\left. {\hat \alpha } \right)\cosh \theta _q  - \left( {\hat \alpha '} \right.|{\bf \tilde v}_q^{\dagger}  |\left. {\hat \alpha } \right)\sinh \theta _q  = 0,\; \forall q, \;\alpha  \ne \alpha ', \; \left( {\hat \alpha } \right.|\left. {\hat \gamma } \right) \ne 0\;,
\end{equation}

\begin{equation}
\left( {\hat \alpha '} \right.|{\bf w}_q^{\dagger}  |\left. {\hat \alpha } \right)\sinh \theta _q  - \left( {\hat \alpha '} \right.|{\bf \tilde v}_q |\left. {\hat \alpha } \right)\cosh \theta _q  = 0,  \;\forall q, \;\alpha  \ne \alpha ',\;\left( {\hat \alpha } \right.|\left. {\hat \gamma } \right) \ne 0\;,
\end{equation}

\end{mathletters}

\noindent and similarly,

\begin{equation}
\label{eq64}
\left( {\hat \gamma _\alpha  } \right.|{\bf W} - {\bf \tilde K}|\left. {\hat \gamma _\alpha  } \right) = \sum\limits_q {\left( {\beta _{q\alpha }^* {\bf w}_q  + \beta _{q\alpha } {\bf w}_q^{\dagger}  } \right)}  - {\bf \tilde K}_{lat} \;,
\end{equation}

\noindent which, inserted in Eq. (\ref{eq62}), leads to

\begin{equation}
\label{eq65}
i\hbar \left( {\hat \alpha ' \left |{ \frac{d\hat \alpha }{dt} } \right.} \right) = \left( {\hat \alpha '} \right.|{\bf H}_{ex}  - {\bf \tilde K}_{lat}  + \sum\limits_q {\left( {\beta _{q\alpha }^* {\bf w}_q  + \beta _{q\alpha } {\bf w}_q^{\dagger}  } \right)} |\left. {\hat \alpha } \right),  \; \forall \;\alpha  \ne \alpha ',\;\left( {\hat \alpha } \right.|\left. {\hat \gamma } \right) \ne 0\;.
\end{equation}

\noindent As before, since $\hat \alpha '$ spans a complete lattice basis, Eqs.(\ref{eq63}) and (\ref{eq65}) indicate that contributing $\hat \alpha $'s, $\left( {\hat \alpha } \right.|\left. {\hat \gamma } \right) \ne 0$, satisfy operatorial equations of the form

\begin{mathletters}
\label{eq66}

\begin{equation}
\cosh \theta _q \;{\bf w}_q \cdot \hat \alpha  - \sinh \theta _q \;{\bf \tilde v}_q^{\dagger}  \cdot \hat \alpha  = \eta _{q,\alpha } \hat \alpha \;,
\end{equation}

\begin{equation}
\sinh \theta _q \;{\bf w}_q^{\dagger} \cdot \hat \alpha  - \cosh \theta _q \;{\bf \tilde v}_q \cdot \hat \alpha  = \bar \eta _{q,\alpha } \hat \alpha \;,
\end{equation}

\end{mathletters}

\noindent and, respectively,

\begin{equation}
\label{eq67}
i\hbar \frac{{d\hat \alpha }}{{dt}} = \left[ {{\bf H}_{lat}  - {\bf \tilde K}_{lat}  + \sum\limits_q {\left( {\beta _{q\alpha }^* {\bf w}_q  + \beta _{q\alpha } {\bf w}_q^{\dagger}  } \right)}  + \Omega _\alpha  } \right]\hat \alpha \;,
\end{equation}

\noindent where $\eta _{q,\alpha } $, $\bar \eta _{q,\alpha } $ and $\Omega _\alpha  $ are scalars. The explicit expressions for the latter can be identified from the contractions of Eq. (\ref{eq57}) with the basis operators $\hat \alpha ' \cdot |\left. {1_{q',\alpha '} } \right)$, $\hat \alpha ' \cdot |\left. {\tilde 1_{q',\alpha '} } \right)$, and, correspondingly, $\hat \alpha ' \cdot \hat \gamma _{\alpha '} $. A straightforward calculation yields

\begin{mathletters}
\label{eq68}

\begin{equation}
\eta _{q,\alpha }  \equiv \left( {\hat \alpha } \right.|\left( {1_{q,\alpha } } \right.|{\bf W} - {\bf \tilde K}|\left. {\hat \gamma _\alpha  } \right)|\left. {\hat \alpha } \right) = \left( {i\hbar \dot \beta _{q\alpha }  - \hbar \omega _q \beta _{q\alpha } } \right)\cosh \theta _q \;,
\end{equation}

\begin{equation}
\bar \eta _{q,\alpha }  \equiv \left( {\hat \alpha } \right.|\left( {\tilde 1_{q,\alpha } } \right.|{\bf W} - {\bf \tilde K}|\left. {\hat \gamma _\alpha  } \right)|\left. {\hat \alpha } \right) = \left( {i\hbar \dot \beta _{q\alpha }  - \hbar \omega _q \beta _{q\alpha } } \right)^* \sinh \theta _q \;,
\end{equation}

\end{mathletters}

\noindent and

\[
\Omega _\alpha   \equiv i\hbar \left( {\hat \alpha  \left|{ \frac{d\hat \alpha }{dt} }\right. } \right) - \left( {\hat \alpha } \right.|{\bf H}_{lat} |\left. {\hat \alpha } \right) + \left( {\hat \alpha } \right.|\left( {\hat \gamma _\alpha  } \right.|{\bf W} - {\bf \tilde K}|\left. {\hat \gamma _\alpha  } \right)|\left. {\hat \alpha } \right){\rm   = } 
\]

\begin{equation}
\label{eq69}
{\rm  = }\sum\limits_q {\hbar \omega _q \left| {\beta _{q\alpha } } \right|} ^2  - \frac{{i\hbar }}{2}\sum\limits_q {\left( {\dot \beta _{q\alpha } \beta _{q\alpha }^*  - \beta _{q\alpha } \dot \beta _{q\alpha }^* } \right)}  \;.
\end{equation}

\noindent

After substituting expressions (\ref{eq68}), Eqs.(\ref{eq66}) can be rearranged in such a manner as to separate the contributions in $w_q $ from those in $v_q $. To this end, multiply Eq. (\ref{eq66}a) on the right by $\left( {\hat \alpha ^{\dagger}  \cosh \theta _q } \right)$, and the adjoint of Eq. (\ref{eq66}b) on the left by $\left( { - \hat \alpha ^ +  \sinh \theta _q } \right)$, and add to obtain

\[
\cosh ^2 \theta _q {\rm  }w_q  \cdot \left( {\hat \alpha \hat \alpha ^{\dagger}  } \right) - \sinh ^2 \theta _q {\rm  }\left( {\hat \alpha \hat \alpha ^{\dagger}  } \right) \cdot w_q  = \left( {i\hbar \dot \beta _{q\alpha }  - \hbar \omega _q \beta _{q\alpha } } \right)\left( {\hat \alpha \hat \alpha ^{\dagger}  } \right)\;.
\]

\noindent Similarly, multiply Eq. (\ref{eq66}a) on the left by $\left( {\hat \alpha ^{\dagger}  \sinh \theta _q } \right)$, and the adjoint of Eq. (\ref{eq66}b) on the right by $\left( { - \hat \alpha ^{\dagger}  \cosh \theta _q } \right)$, and add again to get

\begin{equation}
\label{eq70}
\cosh ^2 \theta _q {\rm  }v_q  \cdot \left( {\hat \alpha ^{\dagger}  \hat \alpha } \right) - \sinh ^2 \theta _q {\rm  }\left( {\hat \alpha ^{\dagger}  \hat \alpha } \right) \cdot v_q  = 0 \;.
\end{equation}

\noindent If we recall that $\hat \alpha ^{\dagger}  \hat \alpha  = U_\alpha  \hat \alpha \hat \alpha ^{\dagger}  U_\alpha ^{\dagger}  $ for some unitary transformation $U_\alpha$, and introduce the diagonal decompositions $\hat \alpha \hat \alpha ^{\dagger}   = \sum\limits_\kappa  {\left| {\chi _{\kappa ,\alpha } } \right\rangle \nu _{\kappa ,\alpha } \left\langle {\chi _{\kappa ,\alpha } } \right|} $ and $\hat \alpha ^{\dagger}  \hat \alpha  = \sum\limits_\kappa  {\left| {\bar \chi _{\kappa ,\alpha } } \right\rangle \nu _{\kappa ,\alpha } \left\langle {\bar \chi _{\kappa ,\alpha } } \right|} $, with $\left| {\bar \chi _{\kappa ,\alpha } } \right\rangle  = U_\alpha  \left| {\chi _{\kappa ,\alpha } } \right\rangle $, from Eq. (\ref{eq70}b) it follows that 

\begin{equation}
\label{eq71}
\left( {\cosh ^2 \theta _q {\rm  }\nu _{\kappa '{\rm ,}\alpha }  - \sinh ^2 \theta _q {\rm  }\nu _{\kappa {\rm ,}\alpha } } \right)\left\langle {\bar \chi _{\kappa ,\alpha } } \right.\left| {v_q } \right|\left. {\bar \chi _{\kappa ',\alpha } } \right\rangle  = 0 \;,
\end{equation}

\noindent hence, 

\begin{equation}
\label{eq72}
\left\langle {\bar \chi _{\kappa ,\alpha } } \right.\left| {v_q } \right|\left. {\bar \chi _{\kappa ',\alpha } } \right\rangle  = \left\langle {\bar \chi _{\kappa ,\alpha } } \right.\left| {v_q^{\dagger}  } \right|\left. {\bar \chi _{\kappa ',\alpha } } \right\rangle  = 0
\end{equation}

\noindent whenever $\nu _{\kappa ,\alpha }  \ne 0$ and/or $\nu _{\kappa ',\alpha }  \ne 0$. But then 

\begin{equation}
\label{eq73}
\hat \alpha  \cdot v_q  = \hat \alpha  \cdot v_q^{\dagger}   = 0
\end{equation}

\noindent must hold, and Eqs.(\ref{eq66}) are so reduced to the simpler form

\begin{mathletters}
\label{eq74}

\begin{equation}
{\rm w}_q  \cdot \hat \alpha  = \left( {i\hbar \dot \beta _{q\alpha }  - \hbar \omega _q \beta _{q\alpha } } \right)\hat \alpha \;,
\end{equation}

\begin{equation}
{\rm w}_q^{\dagger}   \cdot \hat \alpha  = \left( {i\hbar \dot \beta _{q\alpha }  - \hbar \omega _q \beta _{q\alpha } } \right)^* \hat \alpha \;.
\end{equation}

\end{mathletters}

\noindent Equations (\ref{eq73}) show that all gauge couplings $v_q $ compatible with the ansatz (\ref{eq42}) for $\hat \gamma $ give null contribution to the problem and cannot be employed for an eventual alleviation of computational complexity. Finally, use of the quasi diagonal representation $\hat \alpha  = \sum\limits_\kappa  {\left| {\chi _{\kappa ,\alpha } } \right\rangle \sqrt {\nu _{\kappa ,\alpha } } \left\langle {\bar \chi _{\kappa ,\alpha } } \right|} $ [corresponding to the diagonal representations for $\hat \alpha \hat \alpha ^{\dagger}  $ and $\hat \alpha ^{\dagger}  \hat \alpha $ given above] in Eqs.(\ref{eq74}), leads to 

\begin{mathletters}
\label{eq75}

\begin{equation}
\nu _{\kappa ,\alpha } \left[ {{\rm w}_q \left| {\chi _{\kappa ,\alpha } } \right\rangle  - \left( {i\hbar \dot \beta _{q\alpha }  - \hbar \omega _q \beta _{q\alpha } } \right)\left| {\chi _{\kappa ,\alpha } } \right\rangle } \right] = 0 \;,
\end{equation}

\begin{equation}
\nu _{\kappa ,\alpha } \left[ {{\rm w}_q^{\dagger}  \left| {\chi _{\kappa ,\alpha } } \right\rangle  - \left( {i\hbar \dot \beta _{q\alpha }  - \hbar \omega _q \beta _{q\alpha } } \right)^* \left| {\chi _{\kappa ,\alpha } } \right\rangle } \right] = 0\;,
\end{equation}

\end{mathletters}

\noindent and reveals that the states $\left| {\chi _{\kappa ,\alpha } } \right\rangle $ contributing to $\hat \alpha $ [$\nu _{\kappa ,\alpha }  \ne 0$] must necessarily satisfy the eigenvalue equations

\begin{mathletters}
\label{eq76}

\begin{equation}
{\rm w}_q \left| {\chi _{\kappa ,\alpha } } \right\rangle  = \left( {i\hbar \dot \beta _{q\alpha }  - \hbar \omega _q \beta _{q\alpha } } \right)\left| {\chi _{\kappa ,\alpha } } \right\rangle \;,
\end{equation}

\begin{equation}
{\rm w}_q^{\dagger}  \left| {\chi _{\kappa ,\alpha } } \right\rangle  = \left( {i\hbar \dot \beta _{q\alpha }  - \hbar \omega _q \beta _{q\alpha } } \right)^* \left| {\chi _{\kappa ,\alpha } } \right\rangle \;.
\end{equation}

\end{mathletters}

\noindent Finally, in view of Eqs.(\ref{eq74}) and (\ref{eq59}), the effective evolution Eq. (\ref{eq67}) for $\hat \alpha $ reduces to the unperturbed form

\begin{equation}
\label{eq77}
i\hbar \frac{{d\hat \alpha }}{{dt}} = \left[ {{\bf H}_{lat}  - {\bf \tilde K}_{lat}  - \Omega _\alpha  } \right]\hat \alpha \;.
\end{equation}

\noindent It is immediate that Eq. (\ref{eq77}) preserves the orthogonality of distinct $\hat \alpha $'s, i.e., $d\left( {\hat \alpha '|\hat \alpha } \right)/dt = 0$, as required by the ansatz for $\hat \gamma $.

While the self-consistency constraints for the pure state case assumed the form of modified conditions for decoherence-free propagation, Eqs.(\ref{eq73}) and (\ref{eq74}) show that in the finite-temperature case, self-consistency corresponds exactly to the conditions for decoherence-free decoupling. The reason for this is seen in the effect of the interaction term $\left( {{\bf W} - {\bf \tilde K}} \right)$ on any $D_2$ product, i.e., 

\begin{equation}
\label{eq78}
\left( {{\bf W} - {\bf \tilde K}} \right)\hat \alpha  \cdot \hat \gamma _\alpha   = \hat \alpha  \cdot \left[ {\sum\limits_q {\left( {\mu _{q,\alpha } {\bf b}_q^{\dagger}   + \mu _{q,\alpha }^* {\bf b}_q } \right)} {\rm  }\hat \gamma _\alpha  } \right] - \left[ {{\bf \tilde K}_{lat} \hat \alpha } \right] \cdot \hat \gamma _\alpha\;,
\end{equation}

\noindent which means that the interaction becomes decoupled, regardless of the exact nature of the state for the phonon subsystem. Here, $\mu _{q,\alpha } $ denotes the eigenvalue of $w_q $ corresponding to $ \hat \alpha $  [$\mu _{q,\alpha }  = \eta _{q,\alpha } /\cosh \theta _q $].\\

We may conclude now that a finite-temperature state operator of the form (\ref{eq42}) is an exact solution of the evolution equation (\ref{eq56}) if and only if
 
\noindent a) the lattice-phonon interaction ${\bf W}$ is linear in the phonon degrees of freedom [Eq. (\ref{eq59})] ;

\noindent b) the lattice operators $\hat \alpha $ are state operators on decoherence-free subspaces of the lattice subsystem, i.e., on [degenerate] common eigenstates of $w_q $ and $w_q^{\dagger}  $ [Eqs.(\ref{eq74}) and (\ref{eq76})], which are, simultaneously, solutions of the evolution equation (\ref{eq77});

\noindent c) the phonon displacement parameters $\beta _{q\alpha } $ evolve according to

\begin{equation}
\label{eq79}
i\hbar \dot \beta _{q\alpha }  - \hbar \omega _q \beta _{q\alpha }  = \mu _{q,\alpha }\;.
\end{equation}

\noindent In addition,

\noindent d) compatible gauge generators act on the lattice states only, ${\bf \tilde K}\hat \alpha  = {\bf \tilde K}_{lat} \hat \alpha $ for all $\hat \alpha $, and the lattice energy shift $\Omega _\alpha  $ in Eq. (\ref{eq77}) amounts to [see Eq. (\ref{eq69})]

\begin{equation}
\label{eq80}
\Omega _\alpha  \left( t \right) =  - \frac{1}{2}\sum\limits_q {\left( {\mu _{q\alpha } \beta _{q\alpha }^*  + \mu _{q\alpha }^* \beta _{q\alpha } } \right)} \;.
\end{equation}

Obviously, this result is strongly reminiscent of the pure state case examined in Sec. II, but involves a simpler form of ${\bf W}$ and the additional constraint (\ref{eq76}b) on the lattice states, which calls for exact decoherence free propagation in the lattice subsystem. It should be noted, however, that this constraint vanishes in the zero temperature limit, when $\sinh \theta _q  = 0$, and the pure state case of Sec. II, for linear phonon coupling, can be recovered identically if needed. For the particular situation when the initial phonon displacements are null, $\beta _{q\alpha }  = 0$, and the initial state is a product between a lattice distribution and a phonon thermal state, $\hat \rho \left( 0 \right) = \hat \rho _{lat} \left( 0 \right)\hat \rho _{ph,T} $, one obtains the following interesting theorem, referred to in the introductory section:\\

{\it A system initially in a product state $\hat \rho \left( 0 \right) = \hat \rho _{lat} \left( 0 \right)\hat \rho _{ph,T} $ evolves into a Gaussian, generalized $D_1$ state given by the ansatz (\ref{eq42}) if and only if the interaction is linear in the phonon coordinates and the initial lattice state is a distribution on a direct sum of orthogonal decoherence-free subspaces, provided any exist. }\\

When the eigenvalues $\mu _{q,\alpha } $ are time independent, Eq. (\ref{eq79}) shows that the displacements $\beta _{q\alpha } $ perform simple harmonic oscillations around displaced equilibrium positions, 

\begin{equation}
\label{eq81}
\beta _{q\alpha } \left( t \right) = \left[ {\beta _{q\alpha } \left( 0 \right) + \frac{{\mu _{q,\alpha } }}{{\hbar \omega _q }}} \right]e^{ - i\omega _q t}  - \frac{{\mu _{q,\alpha } }}{{\hbar \omega _q }}\;.
\end{equation}

\noindent In the peculiar case when the initial displacement brings each mode over the displaced equilibrium position, the reduced phonon density-matrix becomes time independent [see Eq. (\ref{eq46}), and take into account that $\left( {\hat \alpha |\hat \alpha } \right) = const.$, according to Eq. (\ref{eq77})]. But the stationary phonon state is not exactly thermal, unless the overall state is a simple $D_2$ product, and the lattice evolution is confined to a single DFS. Furthermore, since all generalized amplitudes $\left( {\gamma _\alpha  } \right.|\left. {\gamma _{\alpha '} } \right)$ are constant in time, the lattice reduced density-matrix $\hat \rho _{lat}  = \sum\limits_{\alpha ,\alpha '} {\left( {\gamma _\alpha  } \right.|\left. {\gamma _{\alpha '} } \right)\hat \alpha '\hat \alpha ^{\dagger}  } $ evolves in an unperturbed fashion, up to the [constant] energy shifts $\Omega _\alpha  $, i.e.,

\[
\hat \rho _{lat} \left( t \right) = \sum\limits_{\alpha ,\alpha '} {\left( {\gamma _\alpha  } \right.|\left. {\gamma _{\alpha '} } \right)\exp \left[ { - \frac{i}{\hbar }\left( {H_{lat}  - \Omega _{\alpha '} } \right)} \right]{\rm  }\hat \alpha '\left( 0 \right)\hat \alpha ^{\dagger}  \left( 0 \right)\exp \left[ {\frac{i}{\hbar }\left( {H_{lat}  - \Omega _\alpha  } \right)} \right]} \;.
\]

As for pure states, the evolution of self-consistent, density-matrix $D_1 $ states (\ref{eq42}) is disentangled into separable evolutions of the component $D_2 $ states $\hat \alpha  \cdot \hat \gamma _\alpha  $, hence no self-consistent model can generate equations of motion coupling distinct $D_2$ contributions. Again, no nonlinearity survives in the equation of motion for the lattice states, and standard Davydov distributions, for which $\hat \alpha  = \varphi _n c_n^{\dagger}  \left| 0 \right\rangle _{lat} \left\langle {\chi _\alpha  } \right|$, for some orthonormal $\left| {\chi _\alpha  } \right\rangle $ [$\left\langle {{\chi _\alpha  }} \mathrel{\left | {\vphantom {{\chi _\alpha  } {\chi _{\alpha '} }}} \right. \kern-\nulldelimiterspace} {{\chi _{\alpha '} }} \right\rangle  = \delta _{\alpha \alpha '} $], can generate, when self-consistent, only stationary lattice configurations [$\left( {\hat \alpha |\hat \alpha } \right) = \left| {\varphi _n } \right|^2  = const.$]. A similar statement applies to multiquanta, on-site $D_1 $ states and their superpositions. Also, in complete analogy to the pure state case, the relation to the $D_2$ Davydov soliton equations becomes apparent when the equations of motion for $\hat \alpha $ and $\beta _{q\alpha } $ are cast in the form

\begin{mathletters}
\label{eq82}

\begin{equation}
i\hbar \frac{{d\hat \alpha }}{{dt}} = \left[ {{\bf H}_{lat}  - {\bf \tilde K}_{lat}  + \left\langle {\gamma _\alpha  } \right.\left| {\bf W} \right|\left. {\gamma _\alpha  } \right\rangle  + \Omega _\alpha  } \right]\hat \alpha \;,
\end{equation}

\begin{equation}
i\hbar \dot \beta _{q\alpha }  - \hbar \omega _q \beta _{q\alpha }  = \left\langle {\hat \alpha } \right.\left| {{\bf w}_q } \right|\left. {\hat \alpha } \right\rangle\;.
\end{equation}

\end{mathletters}

\noindent If $\hat \alpha $ is given a linear parametrization, one can recover Davydov-like equations, but the self-consistency constraints (\ref{eq76}) for $\hat \alpha $ reduce the interaction term in Eq. (\ref{eq82}a) to a scalar contribution. Even this scalar cannot have a nonlinear dependence on $\hat \alpha $, since the right-hand side of Eq. (\ref{eq82}b) becomes independent of $\hat \alpha $.

At last, the self-consistency conditions (\ref{eq76}) can be stated alternatively as the initial state constraint 

\begin{equation}
\label{eq83}
\bf{\Lambda }_\alpha  \hat \alpha \left( 0 \right) = 0\;,
\end{equation}

\noindent with

\[
 \bf{\Lambda }_\alpha   = \int\limits_0^\infty  {dt\sum\limits_q {\left[ {\left( {{\bf w}_{q,\alpha }^{\dagger}  \left( t \right) - \mu _{q,\alpha }^ *  \left( t \right)} \right)\left( {{\bf w}_{q,\alpha } \left( t \right) - \mu _{q,\alpha } \left( t \right)} \right) + } \right.} }
\]

\begin{equation}
\label{eq84}
 + \left. {\left( {{\bf w}_{q,\alpha } \left( t \right) - \mu _{q,\alpha } \left( t \right)} \right)\left( {{\bf w}_{q,\alpha }^{\dagger}  \left( t \right) - \mu _{q,\alpha }^ *  \left( t \right)} \right)} \right] \;, 
\end{equation}

\noindent and 

\begin{equation}
\label{eq85}
{\bf w}_{q,\alpha } \left( t \right) = {\bf U}_\alpha ^{\dagger}  \left( t \right){\bf w}_q \left( t \right){\bf U}_\alpha (t)\;,
\end{equation}

\noindent where ${\bf U}_\alpha ( t )$ is the unitary propagator corresponding to the equation of motion (\ref{eq77}).

Recall that the above results apply to the finite-temperature version of the ansatz (\ref{eq42}). It is interesting to keep in mind that the zero temperature limit also provides a density-matrix generalization of the pure $D_1$ states, for which the acceptable lattice-phonon couplings resume expression (\ref{eq11}), bilinear in phonon coordinates. This allows, conceivably, an extended class of nontrivial models and the problem can be approached in the same manner. Nevertheless, only models with interactions linear in the phonon coordinates admit self-consistent, finite temperature extensions, and thus bear realistic physical significance. One may note that both the original Davydov model \cite{Davydov} and all its studied versions \cite{improved} belong to this class.

\section{Discussion and examples}
\label{Sec5}

For the reasons outlined above, we limit the discussion to models with linear phonon coupling, and extend the self-consistency conditions for {\it pure} lattice states to include the constraint

\begin{equation}
\label{eq86}
w_q^{\dagger}  \left| \alpha  \right\rangle  = \left( {i\hbar \dot \beta _{q\alpha }  - \hbar \omega _q \beta _{q\alpha } } \right)^* \left| \alpha  \right\rangle\;,
\end{equation}

\noindent and place the state $\left| \alpha  \right\rangle $ in a DFS of the lattice. In other words, $\left| \alpha  \right\rangle $ can be a self-consistent $D_2$ factor if and only if it lies in a lattice DFS. The properties of the decoherence-free subspaces have been studied in detail in connection with quantum computation theory, and the interested reader is referred to the available literature \cite{decoherence1,decoherence2}. Suffice it to say that their existence is determined by the symmetry properties of both the unperturbed Hamiltonian and the system-bath interaction, and thus, a proper understanding involves a Lie-algebraic/group-theoretic framework. However, for an expedient assessment of various Davydov-like models, the brief pedestrian characterization given below proves satisfactory. 

Let us start by noting that instead of Eq. (\ref{eq86}), it is sufficient to require that $\left| \alpha  \right\rangle $, which according to Eqs.(\ref{eq22}) and (\ref{eq14}), satisfies $w_q \left| \alpha  \right\rangle  = \left( {i\hbar \dot \beta _{q\alpha }  - \hbar \omega _q \beta _{q\alpha } } \right)\left| \alpha  \right\rangle $, must belong to an eigenspace S of $w_q $ that is also left invariant by $w_q^{\dagger}  $. Indeed, since $w_q $ reduces to the identity on S, up to a scalar factor, if $w_q^{\dagger}  $ leaves this subspace invariant, it also must be proportional to the identity, by the complex conjugate factor, and the adjoint eigenvalue equation (\ref{eq86}) is recovered necessarily. Since $\left| \alpha  \right\rangle $ belongs to a DFS, the equation of motion for $\left| \alpha  \right\rangle $ reduces to the unperturbed form [see Eqs.(\ref{eq16}), (\ref{eq17}), and (\ref{eq19})]

\begin{equation}
\label{eq87}
i\hbar \left| {\dot \alpha } \right\rangle  = \left( {H_{lat}  - \Omega _\alpha  } \right)\left| \alpha  \right\rangle \;.
\end{equation}

\noindent Hence, in all cases that support a smooth extension to finite-temperature conditions, the self-consistent lattice states are effectively decoupled from the phonon dynamics, up to a phonon modulated phase factor. 

In fact, the remarkable phenomenon highlighted by decoherence-free propagation is that coupling to a [nonequilibrium] thermal bath need not necessarily result in decoherence to a mixt, statistical state. In particular, any self-consistent $D_2$ product with a pure lattice factor $\left| \alpha  \right\rangle $ displays a finite temperature density-matrix 

\[
\rho \left( t \right) = \exp \left[ { - \frac{i}{\hbar }H_{lat} t} \right]\left| {\alpha \left( 0 \right)} \right\rangle \rho _{ph} \left( {\left\{ {\beta _{q \alpha} } \right\},T} \right)\left\langle {\alpha \left( 0 \right)} \right|\exp \left[ {\frac{i}{\hbar }H_{lat} t} \right]\;,
\]

\noindent with

\[
\rho _{ph} \left( {\left\{ {\beta _{q \alpha} } \right\},T} \right) = \prod\limits_q {\frac{1}{{Z_q }}\exp \left[ { - \frac{{\hbar \omega _q }}{{k_B T}}\left[ {b_q^{\dagger}   - \beta _{q \alpha}^* \left( t \right)} \right]\left[ {b_q  - \beta _{q \alpha} \left( t \right)} \right]} \right]} \;.
\]

\noindent It obviously maintains the pure character of the lattice state throughout the evolution, provided the time-dependence of the displacement parameters complies with the self-consistency conditions. For a simple concrete example, let the lattice system be coupled to the phonon bath in a site-homogeneous manner, such that, e.g., $w_q  = \chi _q \sum\limits_n {c_n^{\dagger}  c_n }  = \chi _q \hat N_{lat} $, and let the unperturbed lattice Hamiltonian conserve the number of lattice excitations, i.e., $\left[ {w_q ,{\rm  }\hat N_{lat} } \right] = \left[ {w_q^{\dagger}  ,{\rm  }\hat N_{lat} } \right] = \left[ {H_{lat} ,{\rm  }\hat N_{lat} } \right] = 0$. Then any lattice state with a well-defined number of excited quanta qualifies as a $D_2$ factor state, and for $\hat N_{lat} \left| \alpha  \right\rangle  = \nu \left| \alpha  \right\rangle $, the associated equations of motion for the displacement parameters read $i\hbar \dot \beta _q  - \hbar \omega _q \beta _q  = \nu \chi _q $, with the trivial solution $\beta _{q \alpha} \left( t \right) = \left[ {\beta _{q \alpha} \left( 0 \right) + (\nu \chi _q / \hbar \omega _q )} \right]\exp( - i\omega _q t)  - (\nu \chi _q / \hbar \omega _q )$. Hence, although the lattice interacts with a [thermal] bath, its state is propagated according to the unperturbed dynamics, and if the original state is a pure state, its condition is preserved in time. On the other hand, the bath evolves into a superpositon of Gaussian states, even when initially in a thermal equilibrium state [$\beta_{q \alpha}(0)=0$]. Only if at the outset the bath modes are displaced directly over the displaced equilibrium positions  $\left( { - \nu \chi _q / \hbar \omega _q } \right)$ , does the bath remain in a stationary state. The effect is absolutely robust under variations of the interaction strength, i.e., stronger coupling cannot induce decoherence.

Returning now to the formal characterization of decoherence-free propagation, observe that if the lattice couplings $w_q $ and $w_q^{\dagger}  $ are time independent, $\left| \alpha  \right\rangle $ must dwell in a time-invariant subspace [corresponding to time independent eigenvalues $\mu _{q\alpha } $] and, since $H_{lat} $ alone drives the evolution of $\left| \alpha  \right\rangle $, this subspace must also be an invariant subspace of $H_{lat} $ [the latter can be time dependent]. To see this, let us assume that the eigenvalue $\mu _{q\alpha } $ corresponding to $\left| {\alpha \left( t \right)} \right\rangle $ changes parametrically in time, such that the time derivative of $w_q \left| \alpha  \right\rangle  = \mu _{q\alpha } \left| \alpha  \right\rangle $ reads $w_q \left| {\dot \alpha } \right\rangle  = \mu _{q\alpha } \left| {\dot \alpha } \right\rangle  + (d\mu _{q\alpha}/dt) \left| \alpha  \right\rangle $. If $\left| {\alpha _ \bot  } \right\rangle $ denotes the component of $\left| {\dot \alpha } \right\rangle $ which is orthogonal to all eigenvectors of $w_q $ [$w_q^{\dagger}  $] for the eigenvalue $\mu _{q\alpha } $ [$\mu _{q\alpha }^* $], then $w_q \left| {\alpha _ \bot  } \right\rangle  = \mu _{q\alpha } \left| {\alpha _ \bot  } \right\rangle  + (d\mu _{q\alpha }/dt) \left| \alpha  \right\rangle $. But the latter equation implies that $\left\langle \alpha  \right|w_q \left| {\alpha _ \bot  } \right\rangle  = (d\mu _{q\alpha }/dt) \left\langle {\alpha } \mathrel{\left | {\vphantom {\alpha  \alpha }} \right. \kern-\nulldelimiterspace} {\alpha } \right\rangle  = 0$ [since $\left\langle \alpha  \right|w_q  = \mu _{q\alpha } \left\langle \alpha  \right|$], and thus $(d\mu _{q\alpha }/dt) = 0$. In fact, the requirement that self-consistent states $\left| \alpha  \right\rangle $ must lie in an invariant subspace of the lattice Hamiltonian extends to all cases where the couplings $w_q $ have time-invariant eigenspaces, but not necessarily time independent eigenvalues [e.g., when $w_q  = \sum\limits_n {\chi _{qn} \left( t \right)c_n^{\dagger}  c_n } $]. Applying the same reasoning, it follows that in such situations the time-dependence of the eigenvalues corresponding to a self-consistent $\left| \alpha  \right\rangle $ is limited to the intrinsic time-dependence implied by the form of coupling, i.e., $(d\mu _{q\alpha }/dt) =(\partial \mu _{q\alpha } / \partial t)$. In the most restrictive situation, when $\left| \alpha  \right\rangle$  is a nondegenerate eigenvector for at least one $w_q $ [$w_q^{\dagger}  $], then it can be solution of Eq. (\ref{eq87}) if and only if it is also a time independent eigenstate of $H_{lat} $. 

The necessary and sufficient condition that a common eigenspace S of all $w_q $ and $w_q^{\dagger}  $ be a decoherence free subspace, i.e., that it remain a common eigenspace of the $w_q $'s and $w_q^{\dagger}  $'s under the dynamics driven by the unperturbed lattice Hamiltonian, is that every $\left| \alpha  \right\rangle $ in S must also satisfy 

\begin{mathletters}
\label{eq88}

\begin{equation}
\left( {\frac{i}{\hbar }\left[ {H_{lat} ,{\rm  }w_q } \right] + \frac{{\partial w_q }}{{\partial t}}} \right)\left| {\alpha \left( t \right)} \right\rangle  = \frac{{d\mu _{q\alpha } }}{{dt}}\left| {\alpha \left( t \right)} \right\rangle \;,
\end{equation}

\begin{equation}
\left( {\frac{i}{\hbar }\left[ {H_{lat} ,{\rm  }w_q^{\dagger}  } \right] + \frac{{\partial w_q^{\dagger}  }}{{\partial t}}} \right)\left| {\alpha \left( t \right)} \right\rangle  = \frac{{d\mu _{q\alpha }^* }}{{dt}}\left| {\alpha \left( t \right)} \right\rangle \;,
\end{equation}

\end{mathletters}

\noindent for every $q$, at all times. The proof is trivial. Let $\left| {\alpha \left( t \right)} \right\rangle $ evolve according to Eq. (\ref{eq87}) and satisfy $w_q \left( t \right)\left| {\alpha \left( t \right)} \right\rangle  = \mu _{q,\alpha } \left( t \right)\left| {\alpha \left( t \right)} \right\rangle $ at some instant t. If $w_q \left( {t + \Delta t} \right)\left| {\alpha \left( {t + \Delta t} \right)} \right\rangle  = \mu _{q,\alpha } \left( {t + \Delta t} \right)\left| {\alpha \left( {t + \Delta t} \right)} \right\rangle $ also holds, then one has, to first order in $\Delta t$,

\[
w_q \left( t \right)\left| {\alpha \left( t \right)} \right\rangle  + \Delta t\left[ {w_q \left( t \right)\left| {\dot \alpha \left( t \right)} \right\rangle  + \frac{{\partial w_q }}{{\partial t}}\left| {\alpha \left( t \right)} \right\rangle } \right] = 
\]

\[
=\mu _{q,\alpha } \left( t \right)\left| {\alpha \left( t \right)} \right\rangle  + \Delta t\left[ {\mu _{q,\alpha } \left( t \right)\left| {\dot \alpha \left( t \right)} \right\rangle  + \frac{{d\mu _{q,\alpha } }}{{dt}}\left| {\alpha \left( t \right)} \right\rangle } \right] \;,
\]

\noindent which in view of Eq. (\ref{eq87}) yields Eq. (\ref{eq88}a). Conversely, if $\left| {\alpha \left( t \right)} \right\rangle $ also satisfies Eq. (\ref{eq88}a) at the same instant t, then a slight rearrangement of terms gives  

\[
\frac{d}{{dt}}\left[ {w_q \left( t \right)\left| {\alpha \left( t \right)} \right\rangle } \right] = \frac{d}{{dt}}\left[ {\mu _{q,\alpha } \left( t \right)\left| {\alpha \left( t \right)} \right\rangle } \right] \;,
\]

\noindent and $w_q \left( {t + \Delta t} \right)\left| {\alpha \left( {t + \Delta t} \right)} \right\rangle  = \mu _{q,\alpha } \left( {t + \Delta t} \right)\left| {\alpha \left( {t + \Delta t} \right)} \right\rangle $ follows necessarily. A similar reasoning applied to $w_q^{\dagger}  $ verifies condition Eq. (\ref{eq88}b). If S is time-invariant, then the restriction of $w_q $ to S changes in time only by a time-dependent scalar multiplication, $w_q \left( t \right)|_S  = \lambda \left( t \right)w_q \left( 0 \right)|_S $, and 

\[
\frac{{\partial w_q }}{{\partial t}}\left| {\alpha \left( t \right)} \right\rangle  = \frac{{d\mu _{q\alpha } }}{{dt}}\left| {\alpha \left( t \right)} \right\rangle \;,
\]

\noindent such that we obtain the following restricted criterion:\\

{\it A time-invariant common eigenspace S of all $w_q $ is a DFS, i.e., is left invariant by both $H_{lat} $ and every $w_q^{\dagger}  $, if and only if it is contained in the common kernel of all commutators $\left[ {w_q ,{\rm  }w_q^{\dagger}  } \right]$ and $\left[ {H_{lat} ,{\rm  }w_q } \right]$. In that case S is also in the kernel of every $\left[ {H_{lat} ,{\rm  }w_q^{\dagger}  } \right]$} [i.e., $\left[ {w_q ,{\rm  }w_{q'}^{\dagger}  } \right]\left| \alpha  \right\rangle  = \left[ {H_{lat} ,{\rm  }w_q } \right]\left| \alpha  \right\rangle  = \left[ {H_{lat} ,{\rm  }w_q^{\dagger}  } \right]\left| \alpha  \right\rangle  = 0$ for every q, and every $\left| \alpha  \right\rangle $ in S]. \\

The above can be given a proof independent of Eqs.(\ref{eq88}). Indeed, assuming that S is time independent, if $w_q^{\dagger}  $ and $H_{lat} $ leave S invariant, for any $\left| \alpha  \right\rangle $ in S it is true that $w_q^ +  w_q \left| \alpha  \right\rangle  \equiv \mu _{q\alpha } w_q^{\dagger}  \left| \alpha  \right\rangle  = w_q w_q^{\dagger}  \left| \alpha  \right\rangle $ and similarly, $H_{lat} w_q \left| \alpha  \right\rangle  \equiv \mu _{q\alpha } H_{lat} \left| \alpha  \right\rangle  = w_q H_{lat} \left| \alpha  \right\rangle $. It is also true that, for all $\left| \alpha  \right\rangle $ in S, $\left\langle \beta  \right.\left| {w_q^{\dagger}  } \right|\left. \alpha  \right\rangle  = 0$ whenever $\left\langle {\beta } \mathrel{\left | {\vphantom {\beta  \alpha }} \right. \kern-\nulldelimiterspace}{\alpha } \right\rangle  = 0$, and $\left\langle {\alpha '} \right.\left| {w_q^{\dagger}  } \right|\left. \alpha  \right\rangle  = \mu _{q\alpha }^* \left\langle {{\alpha '}} \mathrel{\left | {\vphantom {{\alpha '} \alpha}} \right. \kern-\nulldelimiterspace} {\alpha } \right\rangle $ for all $\left| \alpha  \right\rangle $ and $\left| {\alpha '} \right\rangle $ in S. Hence $w_q^{\dagger}  \left| \alpha  \right\rangle  = \mu _{q\alpha }^* \left| \alpha  \right\rangle $ for any $\left| \alpha  \right\rangle $ in S. But then $H_{lat} w_q^{\dagger}  \left| \alpha  \right\rangle  \equiv \mu _{q\alpha }^* H_{lat} \left| \alpha  \right\rangle  = w_q^{\dagger}  H_{lat} \left| \alpha  \right\rangle $, because $H_{lat} \left| \alpha  \right\rangle $ is necessarily in S. Conversely, if $\left[ {w_q ,{\rm  }w_q^{\dagger}  } \right]\left| \alpha  \right\rangle  = 0$ for all $\left| \alpha  \right\rangle $ in S, then $w_q^{\dagger}  \left| \alpha  \right\rangle $ is in S, $\left\langle {\alpha '} \right.\left| {w_q^{\dagger}  } \right|\left. \alpha  \right\rangle  = \mu _{q\alpha }^* \left\langle {{\alpha '}} \mathrel{\left | {\vphantom {{\alpha '} \alpha }} \right. \kern-\nulldelimiterspace}{\alpha } \right\rangle $ for all $\left| {\alpha '} \right\rangle $ in S and $w_q^{\dagger}  \left| \alpha  \right\rangle  = \mu _{q\alpha }^* \left| \alpha  \right\rangle $. And if $\left[ {H_{lat} ,{\rm  }w_q } \right]\left| \alpha  \right\rangle  = 0$ for any $\left| \alpha  \right\rangle $ in S, then $H_{lat} \left| \alpha  \right\rangle $ is an eigenvector of $w_q $ for the eigenvalue $\mu _{q\alpha } $, and must be in S. It is also immediate that $\left[ {H_{lat} ,{\rm  }w_q^{\dagger}  } \right]\left| \alpha  \right\rangle  = 0$ holds too. 

Virtually all studied versions of the Davydov model display time independent interactions, and therefore fall under the incidence of this prescription for self-consistency. For example, the general Frohlich Hamiltonian 

\begin{equation}
\label{eq89}
H = \sum\limits_{mn} {J_{mn} c_m^{\dagger}  c_n }  + \sum\limits_q {\hbar \omega _q b_q^{\dagger}  b_q }  + \sum\limits_{qn} {\hbar \omega _q \left( {\chi _{qn} b_q^{\dagger}   + \chi _{qn}^* b_q } \right)c_n^{\dagger}  c_n } 
\end{equation}

\noindent employs the couplings

\begin{equation}
\label{eq90}
w_q  = \hbar \omega _q \sum\limits_n {\chi _{qn} c_n^{\dagger}  c_n } \;,\; \forall q,
\end{equation}

\noindent which obviously satisfy $\left[ {w_q ,{\rm  }w_q^{\dagger}  } \right] = 0$ and have as eigenvectors the unperturbed on-site Fock states, for the discrete eigenvalues $\mu _{qn}  = \hbar \omega _q \chi _{qn} $. If all $\chi _{qn} $ are nondegenerate for at least some q, as happens in the proper Davydov model, where $\chi _{qn}  = \chi _q e^{inaq} $, the lattice Hamiltonians admitting self-consistent $D_2$ states under the interactions (\ref{eq90}) can only be of the diagonal form $H_{lat}  = \sum\limits_n {\varepsilon _n c_n^{\dagger}  c_n } $. This is, of course, the well-known result that the $D_1$ states are exact under the phonon-lattice interaction (\ref{eq90}) just in the limit of vanishing hopping between monomers. Only when the $\chi _{qn} $ collapse to degenerate values, e.g., in the low-frequency limit, when $qa \to 0$, can nontrivial $D_2$ states appear for the Hamiltonian (\ref{eq89}). But in order to recover a nontrivial dynamics in the low-frequency limit, the phonon modes must belong to an optical band, whereas the Davydov model uses specifically acoustic modes.  

The situation does not improve when hopping terms are added to expression (\ref{eq90}), to obtain 

\[
w_q  = \hbar \omega _q \sum\limits_n {\left[ {\chi _{qn}^{(1)} c_n^{\dagger}  c_n  + \chi _{qn}^{(2)} \left( {c_{n + 1}^{\dagger}  c_n  + c_n^{\dagger}  c_{n + 1} } \right)} \right]}  
\]

\begin{equation}
\label{eq91}
= \hbar \omega _q \sum\limits_n {\left[ {\chi _q^{(1)} c_n^{\dagger}  c_n  + \chi _q^{(2)} \left( {c_{n + 1}^{\dagger}  c_n  + c_n^{\dagger}  c_{n + 1} } \right)} \right]e^{inaq} }\;\;\forall q \;,
\end{equation}

\noindent as done by Takeno \cite{takeno}, Todorovic {\it et al.} \cite{tosic},  Bartnik {\it et al.} \cite{tuszinski} or Pang \cite{pang}. To the contrary, in this case $w_q $ and $w_q^{\dagger}  $ no longer commute, and do not share eigenvectors, which compromises self-consistency from the outset. For instance, with the translation invariant ansatz in the second line above, one obtains

\begin{equation}
\label{eq92}
\left[ {w_q ,{\rm  }w_q^{\dagger}  } \right] = i\kappa _q^{(1)} \sum\limits_n {\left( {c_n^{\dagger}  c_{n + 1}  - c_{n + 1}^{\dagger}  c_n } \right) + i\kappa _q^{(2)} \sum\limits_n {\left( {c_{n - 1}^{\dagger}  c_{n + 1}  - c_{n + 1}^{\dagger}  c_{n - 1} } \right)} } 
\end{equation}

\noindent with $\kappa _q^{(1)}  = 2{\mathop{\rm Im}\nolimits} \left[ {\chi _q^{(1)} \left( {\chi _q^{(2)} } \right)^* \left( {1 - e^{iqa} } \right)} \right]$ and $\kappa _q^{(2)}  =  - 2\left| {\chi _q^{(2)} } \right|^2 \sin \left( {qa} \right)$. But although $\left[ {w_q ,{\rm  }w_q^{\dagger}  } \right]$ is diagonalized by the simple Fourier transformation $c_n  = \frac{1}{{\sqrt N }}\sum\limits_Q {c_Q e^{iQna} } $ and can have null eigenvalues, none of its kernel states is an eigenstate of $w_q $ or $w_q^{\dagger}  $. As for the proper Davydov model, self-consistent $D_2$ states can only appear in the low-frequency limit, provided the phonons belong to optical modes. 

In view of the above, a search for improved wave functions has no chance of recovering a self-consistent model as long as the structure of the Hamiltonian is not adjusted to support decoherence-free $D_2$ states. The partially dressed state, introduced by Brown and Ivic \cite{Brown-Ivic}, which belongs yet to the standard $D_1$ class, can yield a better approximation for the exact dynamics of an initial $D_1$ configuration under the Davydov Hamiltonian, but does not make any progress toward a self-consistent model. The same holds for the coherent state used by Wang {\it et al.} \cite{WBL} in their vibron soliton model, and for Pang's quasicoherent wave function $\left| \Phi  \right\rangle  = \frac{1}{\lambda }\left[ {1 + \sum\limits_n {\varphi _n c_n^{\dagger}   + \frac{1}{{2!}}\left( {\sum\limits_n {\varphi _n c_n^{\dagger}  } } \right)^2 } } \right]\left| 0 \right\rangle _{lat} $ \cite{pang}. In these cases it suffices, in fact, to note that the interactions (\ref{eq90}) and (\ref{eq91}) commute with the lattice number operator, $\hat N_{lat}  = \sum\limits_n {c_n^{\dagger}  c_n } $, hence, any eigenvectors are also eigenvectors of $\hat N_{lat} $. But a coherent or quasicoherent state is not an eigenvector of $\hat N_{lat} $, and cannot generate a self-consistent $D_2$ state even in the low-frequency limit. Yet the underlying idea, that the lattice state $\left| \alpha  \right\rangle $ could be a coherent state [or an approximation thereof], does provide a good starting point for the construction of a nontrivial self-consistent example.

For instance, consider the widely used version of the [time independent] Frohlich Hamiltonian for a one-dimensional lattice of bosonic oscillators with nearest-neighbor hopping,

\begin{equation}
\label{eq93}
H  = \sum\limits_{n} {\varepsilon c_n^{\dagger}  c_n }  - J\sum\limits_n {\left( {c_{n + 1}^{\dagger}  c_n  + c_n^{\dagger}  c_{n + 1} } \right)}  + \sum\limits_q {\hbar \omega _q b_q^{\dagger}  b_q }  + \sum\limits_{qn} {\hbar \omega _q \left( {\chi _{qn} b_q^{\dagger}   + \chi _{qn}^* b_q } \right)c_n^{\dagger}  c_n } \;,
\end{equation}

\noindent to which we add an external pumping term of the form

\begin{equation}
\label{eq94}
W_{pump}  = \sum\limits_{nq} {\hbar \omega _q \left( {\chi _{qn} b_q^{\dagger}   + \chi _{qn}^* b_q } \right)\left( {\zeta _n \left( t \right)c_n^{\dagger}   + \zeta _n^* \left( t \right)c_n } \right)} \;.
\end{equation}

\noindent In such a case, the corresponding couplings $w_q $ are readily identified as 

\begin{equation}
\label{eq95}
w_q  = \hbar \omega _q \left[ {\sum\limits_n {\chi _{qn} \left( {c_n^{\dagger}   + \zeta _n^* \left( t \right)} \right)\left( {c_n  + \zeta _n \left( t \right)} \right) - \sum\limits_n {\chi _{qn} \left| {\zeta _n \left( t \right)} \right|^2 } } } \right] \;,
\end{equation}

\noindent satisfy $\left[ {w_q ,w_q^{\dagger}  } \right] = 0$, and have obvious time-dependent eigenstates. Let us consider  the ground state

\begin{equation}
\label{eq96}
\left| {\alpha \left( t \right)} \right\rangle  = e^{i{\rm  }\Theta _\alpha  \left( t \right)} \exp \left[ { - \sum\limits_n {\left( {\zeta _n \left( t \right)c_n^{\dagger}   - \zeta _n^* \left( t \right)c_n } \right)} } \right]\left| 0 \right\rangle _{lat} \;.
\end{equation}

\noindent Since $\left| \alpha  \right\rangle $ must be simultaneously a solution of the unperturbed problem

\begin{equation}
\label{eq97}
i\hbar \frac{{d\left| \alpha  \right\rangle }}{{dt}} = \left[ {\sum\limits_{mn} {\varepsilon c_n^{\dagger}  c_n }  - J\sum\limits_n {\left( {c_{n + 1}^{\dagger}  c_n  + c_n^{\dagger}  c_{n + 1} } \right)}  - \Omega _\alpha  } \right]\left| \alpha  \right\rangle \;, 
\end{equation}

\noindent it is necessarily a product of coherent states for the lattice normal modes , with a phase factor  

\begin{equation}
\label{eq98}
\Theta _\alpha  \left( t \right) = \frac{i}{\hbar }\int_0^t {d\tau {\rm  }\Omega _\alpha  \left( \tau  \right)}\;.
\end{equation}

\noindent Hence the pump-induced displacements must amount to $\zeta _n \left( t \right) = \sum\limits_q {\zeta _q \left( 0 \right)e^{inqa} e^{ - i\Omega _q t} } $, and satisfy

\begin{equation}
\label{eq99}
i\hbar \frac{{d\zeta _n }}{{dt}} = \varepsilon \zeta _n  - J\left( {\zeta _{n - 1}  + \zeta _{n + 1} } \right) \;.
\end{equation}

\noindent where the lattice frequencies read $\hbar \Omega _q  = \varepsilon  - 2J\cos \left( {qa} \right)$. The corresponding equations of motion for the phonon displacements are now

\begin{equation}
\label{eq100}
i\hbar \frac{{d\beta _{q\alpha } }}{{dt}} = \hbar \omega _q \left[ {\beta _{q\alpha }  - \sum\limits_n {\chi _{qn} \left| {\zeta _n \left( t \right)} \right|^2 } } \right] \;,
\end{equation}

\noindent and apply at both zero and finite-temperature. This is a simple example showing that external pumping can be instrumental in maintaining the lattice in a pure coherent state at arbitrary temperature. Moreover, it also verifies that a Frohlich-like pumping effect can be produced, albeit on the bath modes, due to the nonlinearity in Eq. (\ref{eq100}). Indeed, let us assume, as in the Davydov model, that the bath phonons belong to an accoustic branch of $N$ distinct modes, such that $\chi_{qn}= \chi_q e^{-iqna}$ and $\omega_q=\omega_{-q}$. Taking advantage of the degeneracy of the lattice spectrum, let the external pump act at the lattice frequency $\Omega_Q$, such that $\zeta_n = (\zeta_+ e^{iQna} + \zeta_- e^{-iQna})e^{-i\Omega_Q t}$. But then the driving term for the phonon displacements in Eq. (\ref{eq100}) amounts to 

\[
\chi _q \sum\limits_n {e^{ - iqna} \left| {\zeta _n \left( t \right)} \right|} ^2 = N \chi _q \left[ {\left( {\left| {\zeta _ +  } \right|^2  + \left| {\zeta _ -  } \right|^2 } \right)\delta _{q,0}  + \zeta _ +  \zeta _ - ^* \delta _{q,2Q}  + \zeta _ + ^* \zeta _ -  \delta _{q, - 2Q} } \right]
\]

\noindent which means that only the fundamental bath mode and the modes of frequency $\omega_{2Q}$ experience time independent external driving. Therefore, if the phonon bath is initially in thermal equilibrium, all modes that are not affected by the external pump remain in thermal equilibrium, while the three modes $q = 0, \pm 2Q$ are each driven in a Gaussian state, eventually of macroscopic displacement [see the presence of the $N$ factor above]. In another limit case, when the initial displacements of the bath modes coincide with the displaced equilibrium positions set by the amplitude of the pump, the bath remains in a stationary, nonequilibrium state, while the lattice oscillates harmonically at the pump frequency. Unless $Q=0$ or an accidental degeneracy intervenes, the pumping frequency for the coherent lattice state is distinct from that of the coherently driven modes of the bath, and the process obviously qualifies as a type of Frohlich effect. But contrary to the usual picture, it is the bath that is driven into a nontrivial state, while the primary, lattice subsystem experiences no nonlinearities, and is maintained in a pure state.

For closure, let us illustrate more fundamental features of the coherent-product ansatz [Eq. (\ref{eq42})] in another simple example with exactly tractable decoherence-free propagation. Consider a symmetrical dimer with site-independent Frohlich interaction and phonon-assisted tunneling, described by

\begin{equation}
\label{eq101}
H =  \left( {c_1^{\dagger}  c_1  + c_2^{\dagger}  c_2 } \right)\left[ \varepsilon + \sum\limits_q {\left( {\chi _q b_q^ +   + \chi _q^* b_q } \right)} \right]  -  \left( {c_1^{\dagger}  c_2  + c_2^{\dagger}  c_1 } \right) \left[ J + \sum\limits_q {\left( {\lambda _q b_q^{\dagger}   + \lambda _q^* b_q } \right)}\right]  + \sum\limits_q {\hbar \omega _q b_q^{\dagger}  b_q }  \;.
\end{equation}

\noindent The simple transformation

\begin{equation}
\label{eq102}
c_j  = \frac{1}{{\sqrt 2 }}\left( {\bar c_1  + (-1)^{j-1}\bar c_2 } \right), \;j=1,\;2\;,
\end{equation}

\noindent brings the Hamiltonian (\ref{eq101}) to the evidently solvable form

\begin{equation}
\label{eq103}
H = {\bar \varepsilon} _1 {\bar c}_1^{\dagger}  {\bar c}_1  + {\bar \varepsilon }_2 {\bar c}_2^{\dagger}  {\bar c}_2  + {\bar c}_1^{\dagger}  {\bar c}_1 \sum\limits_q {\left( {\chi _{q,1} b_q^{\dagger}   + \chi _{q,1}^* b_q } \right)}+ {\bar c}_2^{\dagger}  {\bar c}_2 \sum\limits_q {\left( {\chi _{q,2} b_q^{\dagger}   + \chi _{q,2}^* b_q } \right)}  + \sum\limits_q {\hbar \omega _q b_q^{\dagger}  b_q }\;,
\end{equation}

\noindent where the renormalized energies and coupling constants read ${\bar \varepsilon} _1  = \varepsilon  - J$, ${\bar \varepsilon} _2  = \varepsilon  + J$, $\chi _{q,1}  = \chi _q  - \lambda _q $ and, respectively, $\chi _{q,2}  = \chi _q  + \lambda _q $. Since $\left[ {H,{\rm  }\bar c_1^{\dagger}  \bar c_1 } \right] = \left[ {H,{\rm  }\bar c_2^{\dagger}  \bar c_2 } \right] = 0$, the interaction with the bath does not mediate an energy exchange between the monomers, but affects only their coherent correlations through a process of dephasing. In its fermion/two-level realization, this model is easily recognized as equivalent to the exactly solvable Jaynes-Cummings model of quantum optics \cite{Jaynes} and to the Caldeira-Leggett model of quantum dissipation \cite{Leggett}. Various versions have also enjoyed attention lately in studies of decoherence in quantum registers \cite{decoherence3,PSE}. Confining the discussion to single-quantum lattice states, so as to cover both the fermionic and the bosonic case, we apply the ansatz (\ref{eq42}) and construct the self-consistent, statistical superposition

\begin{equation}
\label{eq104}
\hat \gamma  = \varphi _1 \left( t \right)\left[ {\bar c_1^{\dagger}  \left| 0 \right\rangle _{lat} \left\langle {\phi _1 } \right|} \right]\hat \gamma _1  + \varphi _2 \left( t \right)\left[ {\bar c_2^{\dagger}  \left| 0 \right\rangle _{lat} \left\langle {\phi _2 } \right|} \right]\hat \gamma _2 \;,
\end{equation}

\noindent with $\varphi _j \left( t \right) = \varphi _j \left( 0 \right)\exp \left[ { - \frac{i}{\hbar }\left( {\bar \varepsilon _j t - \int_0^t {d\tau {\rm  }\Omega _j \left( \tau  \right)} } \right)} \right]$ scalar amplitudes, $\left| {\phi _j } \right\rangle $ normalized, but not necessarily orthogonal, lattice state vectors, and 

\begin{equation}
\label{eq106}
\hat \gamma _j  = \prod\limits_q {\rm  } \exp \left[ {\beta _{qj} \left( t \right)b_q^{\dagger}   - \beta _{qj}^* \left( t \right)b_q } \right]\frac{1}{{\sqrt {Z_q } }}\exp \left[ { - \frac{{\hbar \omega _q }}{{2k_B T}}b_q^{\dagger}  b_q } \right]\;,
\end{equation}

\noindent where $\beta _{qj} \left( t \right) = \left[ {\beta _{qj} \left( 0 \right) + \frac{\chi _{q,j} }{\hbar \omega _q } } \right]e^{ - i\omega _q t}  - \frac{\chi _{q,j} }{\hbar \omega _q}$. The phonon driven energy shifts $\Omega _j $ are given by Eq. (\ref{eq80})  with the appropriate substitutions. Then, the total density-matrix for the dimer-bath system reads 

\begin{equation}
\label{eq108}
\hat \rho  = \left| {\varphi _1 } \right|^2 \left| {\bar 1} \right\rangle \left\langle {\bar 1} \right|\hat \gamma _1 \hat \gamma _1^{\dagger}   + \left| {\varphi _2 } \right|^2 \left| {\bar 2} \right\rangle \left\langle {\bar 2} \right|\hat \gamma _2 \hat \gamma _2^{\dagger} + \varphi _1 \varphi _2^* \left\langle {{\phi _1 }}
 \mathrel{\left | {\vphantom {{\phi _1 } {\phi _2 }}}
 \right. \kern-\nulldelimiterspace}
 {{\phi _2 }} \right\rangle \left| {\bar 1} \right\rangle \left\langle {\bar 2} \right|\hat \gamma _1 \hat \gamma _2^{\dagger}   + \varphi _1^* \varphi _2 \left\langle {{\phi _2 }}
 \mathrel{\left | {\vphantom {{\phi _2 } {\phi _1 }}}
 \right. \kern-\nulldelimiterspace}
 {{\phi _1 }} \right\rangle \left| {\bar 2} \right\rangle \left\langle {\bar 1} \right|\hat \gamma _2 \hat \gamma _1^{\dagger} \;,
\end{equation}

\noindent where we have denoted, for simplicity, $\left| {\bar 1} \right\rangle  = \bar c_1^{\dagger}  \left| 0 \right\rangle _{lat} $ and $\left| {\bar 2} \right\rangle  = \bar c_2^{\dagger}  \left| 0 \right\rangle _{lat} $, and the reduced density-matrix for the dimer is obtained accordingly as

\begin{equation}
\label{eq109}
\hat \rho _{{\rm dimer}}  \equiv Tr_{ph} \hat \rho  = \left| {\varphi _1 } \right|^2 \left| {\bar 1} \right\rangle \left\langle {\bar 1} \right| + \left| {\varphi _2 } \right|^2 \left| {\bar 2} \right\rangle \left\langle {\bar 2} \right| + \varphi _1 \varphi _2^* \left\langle {{\phi _1 }} \mathrel{\left | {\vphantom {{\phi _1 } {\phi _2 }}} \right. \kern-\nulldelimiterspace}{{\phi _2 }} \right\rangle \left( {\hat \gamma _2 } \right.|\left. {\hat \gamma _1 } \right)\left| {\bar 1} \right\rangle \left\langle {\bar 2} \right| + \varphi _1^* \varphi _2 \left\langle {{\phi _2 }} \mathrel{\left | {\vphantom {{\phi _2 } {\phi _1 }}} \right. \kern-\nulldelimiterspace}{{\phi _1 }} \right\rangle \left( {\hat \gamma _1 } \right.|\left. {\hat \gamma _2 } \right)\left| {\bar 2} \right\rangle \left\langle {\bar 1} \right| \;.
\end{equation}

\noindent Since $\left| {\varphi _j \left( t \right)} \right| = \left| {\varphi _j \left( 0 \right)} \right|$, the occupation probabilities of the states $\left| {\bar 1} \right\rangle $ and $\left| {\bar 2} \right\rangle $ are invariant, and only the correlation $\left( {\hat \rho _{{\rm dimer}} } \right)_{\bar 1\bar 2}  = \left( {\hat \rho _{{\rm dimer}} } \right)_{\bar 2\bar 1}^*  = \varphi _1 \varphi _2^* \left\langle {{\phi _1 }} \mathrel{\left | {\vphantom {{\phi _1 } {\phi _2 }}} \right. \kern-\nulldelimiterspace} {{\phi _2 }} \right\rangle \left( {\hat \gamma _2 } \right.|\left. {\hat \gamma _1 } \right)$ varies in time, as expected in a typical dephasing process. Of course, the same is not true for the original, localized states $\left| 1 \right\rangle  \equiv c_1^{\dagger}  \left| 0 \right\rangle _{lat}  = \frac{1}{{\sqrt 2 }}\left( {\left| {\bar 2} \right\rangle  - \left| {\bar 1} \right\rangle } \right)$ and $\left| 2 \right\rangle  \equiv c_2^{\dagger}  \left| 0 \right\rangle _{lat}  = \frac{1}{{\sqrt 2 }}\left( {\left| {\bar 1} \right\rangle  + \left| {\bar 2} \right\rangle } \right)$. In terms of the latter, the dimer reduced density-matrix reads

\[
\hat \rho _{{\rm dimer}}  = \left( {\frac{1}{2} + {\mathop{\rm Re}\nolimits} \left( {\hat \rho _{{\rm dimer}} } \right)_{\bar 1\bar 2} } \right)\left| 1 \right\rangle \left\langle 1 \right| + \left( {\frac{1}{2} - {\mathop{\rm Re}\nolimits} \left( {\hat \rho _{{\rm dimer}} } \right)_{\bar 1\bar 2} } \right)\left| 2 \right\rangle \left\langle 2 \right| +  
\]

\begin{equation}
\label{eq110}
{\rm              } + \left[ {\left( {\frac{1}{2}\left( {\left| {\phi _2 } \right|^2  - \left| {\phi _1 } \right|^2 } \right) + i{\mathop{\rm Im}\nolimits} \left( {\hat \rho _{{\rm dimer}} } \right)_{\bar 1\bar 2} } \right)\left| 1 \right\rangle \left\langle 2 \right| + h.c.} \right] \;,
\end{equation}

\noindent and it becomes apparent that the correlation between the physical dimers disappears if and only if $\left| {\phi _1 } \right|^2  = \left| {\phi _2 } \right|^2  = \frac{1}{2}$ and ${\mathop{\rm Im}\nolimits} \left( {\hat \rho _{{\rm dimer}} } \right)_{\bar 1\bar 2}  = 0$. It is also evident that all nontrivial aspects of the dynamics are carried by the matrix element $\left( {\hat \rho _{{\rm dimer}} } \right)_{\bar 1\bar 2} $, and particularly by the bath factor

\begin{equation}
\label{eq111}
\left( {\hat \gamma _2 } \right.|\left. {\hat \gamma _1 } \right) = \left( {\hat \gamma _T } \right.|\exp \left[ { - \sum\limits_q {\left( {\beta _{q2} \left( t \right){\bf b}_q^{\dagger}   - \beta _{q2}^* \left( t \right){\bf b}_q } \right)} } \right] \cdot\exp \left[ {\sum\limits_q {\left( {\beta _{q1} \left( t \right){\bf b}_q^{\dagger}   - \beta _{q1}^* \left( t \right){\bf b}_q } \right)} } \right]|\left. {\hat \gamma _T } \right) \;.
\end{equation}

\noindent Expression (\ref{eq111}) can be evaluated straightforwardly by using the thermal operators (\ref{eq49}), and amounts to $\left( {\hat \gamma _2 } \right.|\left. {\hat \gamma _1 } \right) = \exp[-i\Phi(t)] \exp \left[ { - \Gamma \left( t \right)} \right]$, where $\Phi(t)=i\sum\limits_{q}\left( \beta_{q1}\beta_{q2}^*-\beta_{q1}^*\beta_{q2}\right)$ and

\begin{equation}
\label{eq112}
\Gamma \left( t \right) = \frac{1}{2}\sum\limits_q {\left| {\beta _{q1}  - \beta _{q2} } \right|^2 \coth \left( {\frac{{\hbar \omega _q }}{{2k_B T}}} \right)} \;.
\end{equation}

Under the traditional assumption of initial thermal equilibrium of the bath, with null original displacements, the initial density-matrix corresponding to the ansatz (\ref{eq108}) factorizes in the usual manner, as $\hat \rho \left( 0 \right) = \hat \rho _{lat} \left( 0 \right) \otimes \hat \rho _{ph,T} $, where

\begin{equation}
\label{eq113}
\hat \rho _{lat} \left( 0 \right) = \left| {\varphi _1 \left( 0 \right)} \right|^2 \left| {\bar 1} \right\rangle \left\langle {\bar 1} \right| + \left| {\varphi _2 \left( 0 \right)} \right|^2 \left| {\bar 2} \right\rangle \left\langle {\bar 2} \right| + \varphi _1 \left( 0 \right)\varphi _2^* \left( 0 \right)\left\langle {{\phi _1 }} \mathrel{\left | {\vphantom {{\phi _1 } {\phi _2 }}} \right. \kern-\nulldelimiterspace}{{\phi _2 }} \right\rangle \left| {\bar 1} \right\rangle \left\langle {\bar 2} \right| + \varphi _1^* \left( 0 \right)\varphi _2 \left( 0 \right)\left\langle {{\phi _2 }} \mathrel{\left | {\vphantom {{\phi _2 } {\phi _1 }}} \right. \kern-\nulldelimiterspace}{{\phi _1 }} \right\rangle \left| {\bar 2} \right\rangle \left\langle {\bar 1} \right| \;,
\end{equation}

\noindent and $\rho _{ph,T}  = \hat \gamma _T \hat \gamma _T^{\dagger}  $. Since at later times the bath displacements read $\beta _{qj} \left( t \right) =  (\chi _{qj} / \hbar \omega _q ) \left( {e^{ - i\omega _q t} -1} \right)$, the relaxation function $\Gamma $ becomes

\begin{equation}
\label{eq114}
\Gamma ^0 \left( t \right) = 2 \sum\limits_q { \left| {\lambda _q } \right|^2 \coth \left( { \frac{{\hbar \omega _q }}{{2 k_B T}}} \right) \frac{{\left( {1 - \cos \omega _q t} \right)}}{{\left( {\hbar \omega _q } \right)^2 }}} \;.
\end{equation}

\noindent The physics behind expression (\ref{eq114}) has been discussed at length in Ref.\cite{PSE}. Depending on the spectral density of the bath modes, the relaxation exponent may or may not have a finite limit as time passes to infinity. If it does not reach a plateau, the correlation $\left( {\hat \rho _{{\rm dimer}} } \right)_{\bar 1\bar 2} $ eventually falls to zero and the dimer density-matrix is invariably driven toward a steady distribution diagonal on the states $\left| {\bar 1} \right\rangle $ and $\left| {\bar 2} \right\rangle $, hence, undergoing total dephasing through an environment-induced superselection process \cite{zurek}. The noteworthy point in the above derivation is that the dimer density-matrix (\ref{eq113}), parametrized by three linearly independent, real parameters [$\left| {\phi _1 } \right|^2  + \left| {\phi _2 } \right|^2  = 1$], may span the entire set of statistical dimer states corresponding to the subspace $\left\{ {\left| {\bar 1} \right\rangle ,{\rm  }\left| {\bar 2} \right\rangle } \right\}$ [or $\left\{ {\left| 1 \right\rangle ,{\rm  }\left| 2 \right\rangle } \right\}$]. In other words, Eq. (\ref{eq108}) gives explicitly the general solution of the Liouville-von Neumann problem for the Hamiltonian (\ref{eq101}), with an initial condition of the type $\hat \rho \left( 0 \right) = \hat \rho _{lat} \left( 0 \right) \otimes \hat \rho _{ph,T} $.

From the point of view of coherent behavior, a special mention is reserved once more for the singular situation when the bath displacements remain stationary in their displaced equilibrium positions, i.e., $\beta _{qj} \left( t \right) =  - (\chi _{qj} / \hbar \omega _q )$. Then the bath is left in a steady superposition of thermal Gaussian states, and the corresponding bath factor (\ref{eq112}) becomes constant in time. If the latter does not vanish, the matrix element $\left( {\hat \rho _{{\rm dimer}} } \right)_{\bar 1\bar 2} $ oscillates as $\varphi _1 \left( t \right)\varphi _2^* \left( t \right) = \varphi _1 \left( 0 \right)\varphi _2^* \left( 0 \right)\exp \left[ { - \frac{i}{\hbar }\Delta \Omega {\rm  }t} \right]$, at the frequency  

\begin{equation}
\label{eq115}
\Delta \Omega  \equiv \Omega _2  - \Omega _1  = \sum\limits_q {\frac{{\chi _q \lambda _q^*  + \chi _q^* \lambda _q }}{{\hbar \omega _q }}} \;,
\end{equation}

\noindent and the dimer is kept in a coherently oscillating [or two-level rotating] state, somewhat reminiscent of a soliton state.

\section{Conclusion}
\label{Sec6}

We have analyzed the self-consistency conditions for a fairly wide class of generalized $D_1$ states, based on the Davydov ansatz for soliton propagation in molecular chains. Our extended ansatz is given in a density-matrix [state operator] form [Eq. (\ref{eq42})], which covers, in proper limit cases, both the pure state and the finite-temperature standard Davydov states. In general, it describes strongly entangled, statistical superpositions of lattice and bath states, with the property that the bath is maintained in a statistical superposition of Gaussian thermal states. We find that the exact propagation of such states amounts to an effective decoupling of the lattice subsystem from its boson bath, i.e., to decoherence-free propagation. Given the specific form of the bath state, the lattice-bath interactions compatible with such a phenomenon are limited to linear or bilinear forms in the bath coordinates, but only interactions linear in the bath degrees of freedom allow self-consistent propagation at finite-temperatures. Unfortunately, in all cases, the equations of motion for the ansatz parameters differ from Davydov's soliton equations by additional constraints, which apparently abolish the characteristic nonlinearities of the latter. Just as Brown's validity theorems \cite{brown}, the result is independent of any internal characteristics of the system, including symmetries, range and strength of interactions, or type of phonon branches. This does not affect the suitability of the ansatz as a variational trial ansatz, and the Davydov model may still be a good approximation in different conditions.

As an interesting byproduct, we are left with a set of nontrivial, exact density-matrix solutions for systems supporting decoherence-free propagation. Although exact results for such systems have been already reported [see, e.g., \cite{Venugopalan,PSE}], the present solutions are given in a closed, explicit operatorial form, and are not limited to the usual equilibrium, separable, initial conditions. Consequently, it becomes possible to probe the dynamics of such systems under nonequilibrium, Gaussian states of the bath. We retrieved a Frohlich effect under decoherence-free conditions facilitated by time-dependent, external pumping, and found that certain nonequilibrium states of the bath can be instrumental in maintaining bath-entangled, decoherence-free states of the driven (lattice) subsystem. We believe that this ansatz may prove useful in illustrating a variety of nontrivial aspects of relaxation and decoherence in quantum systems.

\end{document}